\documentclass[a4paper,11pt]{article}

\usepackage{amsmath,amsthm,amsbsy,amssymb,amsfonts,graphicx,mathrsfs,bm,cite,color,accents}
\usepackage{hyperref,bookmark}
\usepackage[scr=boondoxo]{mathalfa}

\makeatletter
\catcode`\@=11
\@addtoreset{equation}{section}

\makeatother

\usepackage[utf8]{inputenc}

\renewcommand{\thefootnote}{\arabic{footnote}}

\newcommand{\Exp}[1]{\operatorname{e}^{#1}}
\newcommand{\diag}{\operatorname{diag}}

\newcommand{\abs}[1]{\lvert {#1} \rvert}
\newcommand{\rmd}{{\mathrm{d}}}
\newcommand{\rmT}{{\tt T}}
\newcommand{\nn}{\nonumber}
\newcommand{\Lie}{\pounds}
\newcommand{\gLie}{\hat{\pounds}}

\newcommand{\cA}{\mathcal A}\newcommand{\cB}{\mathcal B}
\newcommand{\cC}{\mathcal C}\newcommand{\cD}{\mathcal D}

\newcommand{\cH}{\mathcal H}
\newcommand{\cI}{\mathcal I}\newcommand{\cJ}{\mathcal J}
\newcommand{\cK}{\mathcal K}\newcommand{\cL}{\mathcal L}
\newcommand{\cM}{\mathcal M}
\newcommand{\cP}{\mathcal P}

\newcommand{\SO}{\text{SO}}
\newcommand{\UU}{\text{U}}
\newcommand{\SL}{\text{SL}}

\newcommand{\sfi}{\mathsf{i}}
\newcommand{\sfj}{\mathsf{j}}
\newcommand{\sfk}{\mathsf{k}}
\newcommand{\sfl}{\mathsf{l}}

\newcommand{\KK}{K}
\newcommand{\kk}{k}

\newcommand{\gi}{{\scriptscriptstyle\mathcal{I}}}
\newcommand{\gj}{{\scriptscriptstyle\mathcal{J}}}
\newcommand{\gk}{{\scriptscriptstyle\mathcal{K}}}

\newcommand{\Ba}{\alpha}
\newcommand{\Bb}{\beta}
\newcommand{\Bc}{\gamma}
\newcommand{\Bd}{\delta}

\newcommand{\nM}{\mathfrak{n}}

\allowdisplaybreaks[3]

\begin{document}

\begin{titlepage}
\renewcommand{\thefootnote}{\fnsymbol{footnote}}

\vspace*{1.0cm}

\begin{center}
\centerline{\Large\textbf{Gauged sigma models and exceptional dressing cosets}}%
\end{center}
\vspace{1.0cm}

\centerline{\large
{Yuho Sakatani}%
\footnote{E-mail address: \texttt{yuho@koto.kpu-m.ac.jp}}
\ \ and \ \ 
{Shozo Uehara}%
\footnote{E-mail address: \texttt{uehara@koto.kpu-m.ac.jp}}
}

\vspace{0.2cm}

\begin{center}
{\it Department of Physics, Kyoto Prefectural University of Medicine,}\\
{\it 1-5 Shimogamohangi-cho, Sakyo-ku, Kyoto 606-0823, Japan}\\
\end{center}

\vspace*{2mm}

\begin{abstract}
The Poisson--Lie (PL) $T$-duality is a generalized $T$-duality based on the Lie algebra of the Drinfel'd double. In particular, when we consider the PL $T$-duality of a coset space, the dual space is found to be a generalized coset space, which is called the dressing coset. In this paper, we investigate an extension of the dressing cosets to the $U$-duality setup. We propose the gauged actions for various branes in M-theory and type IIB theory, where the generalized metric is constructed by using the Exceptional Drinfel'd Algebra (EDA) and the gauge algebra is a certain isotropic subalgebra of the EDA. By eliminating the gauge fields, the gauged action reduces to the standard brane action on a certain reduced background, which we call the exceptional dressing coset. We also propose an alternative definition of the exceptional dressing cosets based on Sfetsos's approach and reproduce a known example of non-Abelian $T$-duality in the $U$-duality-covariant formulation.
\end{abstract}

\thispagestyle{empty}
\end{titlepage}

\setcounter{footnote}{0}

\newpage

\tableofcontents

\newpage

\section{Introduction}

The Poisson--Lie (PL) $T$-duality \cite{hep-th:9502122,hep-th:9509095} is a natural generalization of Abelian and non-Abelian $T$-duality. 
This is based on the Lie group of a $2D$-dimensional Drinfel'd double $\cD$, and the physical space $M$ on which strings propagate is constructed as a coset space $M=\cD/\tilde{G}$. 
Here $\tilde{G}$ is a $D$-dimensional subgroup of $\cD$ that is isotropic with respect to the $T$-duality-invariant bilinear form $\langle\cdot,\,\cdot\rangle$. 
In general, there are several inequivalent choices of the subgroup $\tilde{G}$, and we can construct several different physical spaces $M=\cD/\tilde{G}$, $M'=\cD/\tilde{G}'$, $\dotsc$.
The PL $T$-duality is a classical equivalence of string theories defined on the family of physical spaces. 
This equivalence still holds even if we consider a further quotient by an isotropic subgroup $F$ of $\cD$ \cite{hep-th:9602162,hep-th:9903170,hep-th:9904188}. 
Namely, string theories defined on a family of the reduced physical spaces $\check{M}=F\backslash\cD/\tilde{G}$, $\check{M}'=F\backslash\cD/\tilde{G}'$, $\dotsc$, are known to be equivalent (at least at the classical level). 
When $\cD$ is a perfect Drinfel'd double, the physical spaces $M=\cD/\tilde{G}$, $M'=\cD/\tilde{G}'$, $\dotsc$, can be identified with group manifolds of $D$-dimensional (isotropic) Lie groups, $G$, $G'$, $\cdots$, and the reduced physical spaces can be expressed as $\check{M}=F\backslash G$, $\check{M}'=F\backslash G'$, $\dotsc$\,. 
In particular, if $F$ is a subgroup of $G$, $\check{M}=F\backslash G$ is a standard coset space, but in general, they are non-trivial coset spaces, which is called the dressing coset. 
Then, if we perform a PL $T$-duality transformations, a coset space is generally mapped to dressing cosets. 

Recently, a $U$-duality extension of the Drinfel'd double, called the Exceptional Drinfel'd Algebra (EDA), has been proposed in \cite{1911.06320,1911.07833} and various aspects of the EDA have been studied in \cite{2007.08510,2009.04454,2001.09983,2003.06164,2006.12452,2011.11424,2012.13263,2103.01139,2107.00091,2202.00355,2203.01838}. 
Then, it is of interest to investigate whether the notion of the dressing cosets can be generalized to the case of the EDA. 
A difficulty in this generalization arises from the fact that a group-like extension of the EDA has not yet been clarified. 
The EDA $\mathfrak{d}(\mathfrak{e}_{\nM(\nM)})$ is a Leibniz algebra, and unlike the Lie algebra, we do not have a group manifold associated with the EDA. 
When we consider the generalized $U$-duality, we find several decompositions of the EDA into a pair of subalgebras $\mathfrak{d}(\mathfrak{e}_{\nM(\nM)})=\mathfrak{g}\oplus\tilde{\mathfrak{g}}$, where $\mathfrak{g}$ is a maximally isotropic subalgebra and $\tilde{\mathfrak{g}}$ is also a subalgebra. 
Here, $\mathfrak{g}$ is a Lie algebra (due to the isotropy), and we can consider the associated Lie group $G$. 
The group manifold of $G$ has been found to play the role of the physical space $M$ \cite{1911.06320,1911.07833}, and by considering various inequivalent decompositions, $\mathfrak{d}(\mathfrak{e}_{\nM(\nM)})=\mathfrak{g}\oplus\tilde{\mathfrak{g}}=\mathfrak{g}'\oplus\tilde{\mathfrak{g}}'=\cdots$, we can construct various physical spaces $M\simeq G$, $M'\simeq G'$, $\cdots$.
The symmetry which connects these geometries has been known as the generalized $U$-duality. 
A major difference from the $T$-dual case is that we cannot express this physical space $M\simeq G$ as a coset space, such as $\cD/\tilde{G}$. 
In the $T$-dual case, the dressing coset has been defined as a double coset $F\backslash\cD/\tilde{G}$, and $F$ can be an arbitrary isotropic subgroup of $\cD$, but in the $U$-duality generalization, naively, $F\backslash G$ can be defined only when $F$ is a subgroup of $G$. 

In fact, to define a $U$-duality extension of the dressing cosets, it is important to recall the notion of the dressing action. 
For a perfect Drinfel'd double, an arbitrary element $\ell$ of $\cD$ can be factorized into a product $\ell=g\,\tilde{g}$ ($g\in G$, $\tilde{g}\in\tilde{G}$).
Then, using the gauge symmetry associated with the right multiplication by $\tilde{G}$, we can always choose a gauge where $\ell=g\in G$, and the coset space $M=\cD/\tilde{G}$ is identified with a group manifold of $G$. 
Under a left multiplication by an element $f\in F$, $\ell=g$ is mapped to a certain element of $\cD$, and by simultaneously performing a gauge transformation associated with $\tilde{G}$, we can obtain an element $\ell'= g'\in G$. 
The map $g\mapsto g'$, defined in this way, is called the dressing action. 
We can then consider an orbit of the dressing action of $F$ on $G$ even when $F$ is not a subgroup of $G$. 
By regarding this as a gauge orbit, we can construct the dressing coset $F\backslash G$. 
The important point that was noted in \cite{2112.14766} is that the dressing action by $f=1+\epsilon^{\gi}\,T_{\gi}\in F$ ($\epsilon^{\gi}:\text{infinitesimal constant}$) corresponds to a generalized diffeomorphism in double field theory (DFT) \cite{hep-th:9305073,0904.4664,0908.1792,1003.5027,1006.4823} that is generated by the generalized Lie derivative $\epsilon^{\gi}\,\gLie_{\KK_{\gi}}$ along the generalized vector fields $\KK_{\gi}{}^M\equiv \KK_{\gi}{}^A\,E_A{}^M$, which are constructed from the Lie algebra of the Drinfel'd double. 

We here propose that the dressing action in the $U$-duality context is generated by the generalized Lie derivative of exceptional field theory (ExFT) \cite{1008.1763,1111.0459,1206.7045,1208.5884,1308.1673,1312.0614,1312.4542,1406.3348} $\epsilon^{\gi}\,\gLie_{\KK_{\gi}}$ along $\KK_{\gi}{}^I\equiv \KK_{\gi}{}^A\,E_A{}^I(x)$. 
If the gauge group $F$ is a subgroup of $G$, $\KK_{\gi}{}^I$ reduce to the usual right-invariant vector fields on $G$, similar to the case of the Drinfel'd double. 
By requiring that the generalized vector fields $\KK_{\gi}{}^I$ satisfy the generalized Killing equations $\gLie_{\KK_{\gk}}\cM_{IJ}=0$, the brane action turns out to have a global symmetry under the dressing action. 
Then we can promote the global symmetry into a local symmetry and define the gauged sigma model. 
By the construction, this model should describe the motion of a brane on a dressing coset $F\backslash G$, which we call the exceptional dressing coset. 
In principle, eliminating the gauge fields, we can reduce the action to the standard brane action and read off the supergravity fields on the dressing coset. 

To clarify the symmetry of the EDA, it is useful to use the $U$-duality-covariant approach of the brane worldvolume theories. 
As such an approach, here we employ the approach proposed and developed in \cite{1607.04265,1712.10316,2004.09486} (see \cite{1802.00442,1804.07303,1904.06714,2102.00555,2103.03267,2103.08608,2107.10568,2110.13010} for other approaches proposed recently). 
For a $p$-brane in M-theory or type IIB theory, the brane action can be generally expressed as
\begin{align}
 S = \frac{1}{2\,(p+1)} \int_{\Sigma_{p+1}} \Exp{\lambda}\cM_{IJ}(x) \,P^I\wedge *_{\gamma} P^J + \int_{\Sigma_{p+1}} \Omega_{\text{brane}} \,.
\end{align}
The supergravity fields are contained only in the generalized metric $\cM_{IJ}(x)$, and the details of $\Omega_{\text{brane}}$ depend on the brane. 
The worldvolume gauge fields are contained only in $\Omega_{\text{brane}}$. 
At least for the membrane, the M5-branes, and the $(p,q)$-string, the classical equivalence to the standard brane theories has been checked in \cite{1607.04265,1712.10316,2004.09486} and here we use this as the ungauged action. 
We assume the existence of generalized Killing vector fields $\KK_{\gi}{}^I$ ($\gi=1,\dotsc,n$) that form a Lie algebra $\gLie_{\KK_{\gi}}\KK_{\gj}{}^I=-f_{\gi\gj}{}^{\gk}\,\KK_{\gk}{}^I$, which corresponds to the gauge algebra $\mathfrak{f}$ of the dressing action. 
Then the brane action is invariant under the coordinate transformation $\delta x^{\sfi}=\epsilon^{\gi}\,\kk_{\gi}^{\sfi}$ ($\epsilon^{\gi}:\text{constant}$) and we consider promoting the global symmetry into the local symmetry $\delta x^{\sfi}(\sigma)=\epsilon^{\gi}(\sigma)\,\kk_{\gi}^{\sfi}(x(\sigma))$ by introducing the gauge fields $\cA^{\gi}(\sigma)$\,. 
We explain a general strategy to construct the gauged action and construct the gauged actions for various standard branes: the membrane, the M5-brane, the $(p,q)$-string, and the D3-brane. 

After finding the gauged brane actions, we consider eliminating the gauge fields $\cA^{\gi}$. 
We gauge-fix the $n$-dimensional gauge symmetry by setting $n$ coordinates to be constant, and also eliminate the gauge fields $\cA^{\gi}$. 
We then obtain the standard (ungauged) brane action on a $(D-n)$-dimensional exceptional dressing coset. 
Concretely, we consider a $(p,q)$-string action and obtain the metric and the 2-form potential fields, which play the role of the supergravity fields on the exceptional dressing coset. 
We then find a certain issue on the obtained supergravity fields. 
To avoid the issue, we propose an alternative way to construct the supergravity fields on the exceptional dressing cosets, which extends Sfetsos's approach \cite{hep-th:9904188} to the $U$-duality setup. 
This allows us to determine all of the bosonic supergravity fields and makes it possible to discuss the generalized $U$-duality at the level of supergravity equations of motion. 

This paper is organized as follows. 
In section \ref{sec:review-dressing-cosets}, we review the standard dressing cosets and the gauged string sigma model. 
In particular, we explain two equivalent descriptions of the gauged sigma model to make it easy to study the generalization to brane theories. 
In section \ref{sec:review-our-action}, we review the (ungauged) brane actions. 
In section \ref{sec:gauged-brane}, we explain how to construct the gauge-invariant actions, and concretely construct various gauged brane actions in M-theory and type IIB theory. 
In section \ref{sec:RBF}, we discuss the supergravity fields on the exceptional dressing cosets. 
For a $(p,q)$-string, we find that the gauge fields $\cA^{\gi}$ can be generally eliminated and the background fields can be determined from the string action. 
We then discuss a subtle issue and propose an alternative approach based on the Sfetsos's limit. 
As a demonstration of this approach, we discuss non-Abelian $T$-duality of 2-sphere in the $U$-duality setup. 
Section \ref{sec:conclusions} is devoted to conclusion and discussion.

\section{Dressing cosets and the gauged string action}
\label{sec:review-dressing-cosets}

For a classical string on a dressing coset, the gauged sigma models have been studied in \cite{1903.00439,2112.14766}. 
In this section, we review the gauged sigma model of \cite{2112.14766} and stress that there are two equivalent approaches to formulate the gauged sigma model. 

\medskip

\noindent\textbf{1. Group-based approach}\\
The first approach describes the position of the string by using a map $\sigma\mapsto l(\sigma)\in\cD$ from the worldsheet $\Sigma_2$ to an element of a $2D$-dimensional Drinfel'd double $\cD$. 
We introduce gauge fields $\cC(\sigma)$ that are associated with right multiplications by $\tilde{G}$. 
Accordingly, the sigma model describes string theory on the $D$-dimensional coset space $\cD/\tilde{G}$. 
We also introduce gauge fields $\cA(\sigma)$ associated with left multiplications $l(\sigma)\to f(\sigma)\,l(\sigma)$ $(f\in F)$ by an $n$-dimensional subgroup $F$ of $\cD$.
Then, the resulting gauged sigma model corresponds to string theory on the dressing coset $F\backslash \cD/\tilde{G}$. 

\noindent\textbf{2. Coordinate-based approach}\\
In the second approach, the position of the string is described by the embedding function $x^m(\sigma)$, which gives a map from $\Sigma_2$ to the $D$-dimensional coset space $\cD/\tilde{G}$. 
The group element $l(\sigma)$ defined in the first approach can be parameterized as $l(\sigma)=\Exp{\delta^a_m x^m(\sigma)\,T_a}\Exp{\delta_a^m \tilde{x}_m(\sigma)\,T^a}$ where $\{T_a,\,T^a\}$ spans the Lie algebra $\mathfrak{d}$ and $\{T^a\}$ spans $\tilde{\mathfrak{g}}$, which is the Lie algebra of $\tilde{G}$. 
Then one may suspect that we also need to introduce $\tilde{x}_m(\sigma)$.
However, in the gauged action, roughly speaking $\tilde{x}_m$ appears with the combination $\rmd \tilde{x}+\cC(\sigma)$ and we can rename the combination as $P_m(\sigma)$. 
Accordingly, in the second approach, we introduce independent auxiliary fields $P_m(\sigma)$ instead of considering $\tilde{x}_m$ and $\cC$. 
To study the dressing cosets, we again introduce the gauge fields $\cA(\sigma)$ in the same way as the first approach. 
A difference from the first approach is that (infinitesimal) left multiplications are expressed as coordinate transformations along a certain Killing vector field $x^m(\sigma)\to x^m(\sigma) + \epsilon^{\gi}(\sigma)\,\kk_{\gi}^m(x(\sigma))$. 
This transformation looks different from that of the first approach, but these are actually the same transformation.
The gauged action is indeed gauge invariant in both approaches. 

\medskip

When we consider the exceptional dressing cosets, it is not easy to take the first approach because it is not clear how to define the group-like extension of the EDA. 
Accordingly, here we explain the second approach in more detail, and then study its extensions in section \ref{sec:gauged-brane}. 

\subsection{Drinfel'd double}

Here we shortly introduce several notations associated with the Drinfel'd double. 
The Lie algebra of the $2D$-dimensional Drinfel'd double is given by
\begin{align}
 [T_a,\,T_b]=f_{ab}{}^c\,T_c\,,\qquad
 [T^a,\,T^b]=f_c{}^{ab}\,T^c\,,\qquad
 [T_a,\,T^b]= f_a{}^{bc}\,T_c - f_{ac}{}^b\,T^c\,,
\end{align}
where $a=1,\dotsc,D$.
The two subalgebras, generated by $\{T_a\}$ and $\{T^a\}$, are respectively denoted as $\mathfrak{g}$ and $\tilde{\mathfrak{g}}$, and both of these are maximally isotropic with respective to the symmetric bilinear form
\begin{align}
 \langle T_a,\,T_b\rangle = 0 = \langle T^a,\,T^b\rangle\,,\qquad
 \langle T_a,\,T^b\rangle = \delta_a^b\,. 
\end{align}
We shall denote the generators collectively as $\{T_A\}=\{T_a,\,T^a\}$ ($A=1,\dotsc,2D$) and also denote the algebra and the bilinear form as
\begin{align}
 [T_A,\,T_B]=F_{AB}{}^C\,T_C \,,\qquad 
 \langle T_A,\,T_B\rangle =\eta_{AB} \,,\qquad 
 \eta_{AB}\equiv \begin{pmatrix} 0 & \delta_a^b \\ \delta_a^b & 0 \end{pmatrix}.
\end{align}

\subsection{Gauged string action in the first approach}

Let us consider the gauged action \cite{2112.14766}
\begin{align}
 S = \frac{1}{4\pi\alpha'}\int_{\Sigma_2} \Bigl(\frac{1}{2}\,\langle \bm{\cP} \overset{\wedge}{,} *_\gamma \hat{\cH}(\bm{\cP})\rangle + \langle \rho \overset{\wedge}{,}\, l\,\cC\,l^{-1}\rangle 
 - \langle P \overset{\wedge}{,}\, \cA\rangle \Bigr)
 - \frac{1}{4\pi\alpha'}\int_{\cB} \frac{1}{3!}\, \langle \rho\overset{\wedge}{,}\, [\rho,\, \rho]\rangle \,,
\label{eq:DD-string-action}
\end{align}
where $\hat{\cH}:\mathfrak{d}\to\mathfrak{d}$ is a linear map satisfying $\hat{\cH}^2=1$ and $\langle T_A,\,\hat{\cH}(T_B)\rangle=\langle \hat{\cH}(T_A),\,T_B\rangle$, and
\begin{align}
 \rho\equiv \rmd l\,l^{-1}\,,\qquad
 P \equiv \rho + l \,\cC \,l^{-1} \,,\qquad
 \bm{\cP} \equiv P + \cA \,,\qquad
 \partial\cB=\Sigma_2\,.
\label{eq:defs}
\end{align}
The gauge field $\cC=\cC^A\,T_A$ takes value in $\tilde{\mathfrak{g}}$ and can be parameterized as $(\cC^A)=(0,\,\cC_a)$. 
The gauge field $\cA$ is defined to take value in an $n$-dimensional Lie algebra $\mathfrak{f}$ that is an isotropic subalgebra of $\mathfrak{d}$. 
We denote the generators of $\mathfrak{f}$ as
\begin{align}
 T_{\gi} \equiv \KK_{\gi}{}^A\,T_A\,,\qquad [T_{\gi},\,T_{\gj}] = f_{\gi\gj}{}^{\gk}\,T_{\gk}\,,\qquad 
 \langle T_{\gi},\,T_{\gj} \rangle=0\,,
\end{align}
where $\gi=1,\dotsc,n$ and $\KK_{\gi}{}^A$ is constant. 
Then we expand the gauge field as $\cA=\cA^{\gi}\,T_{\gi}$\,. 

We can show that the action \eqref{eq:DD-string-action} has the following two gauge symmetries.
\begin{enumerate}
\item Under the right multiplication by an element of $\tilde{G}$, $l(\sigma)\to l(\sigma)\,\tilde{h}(\sigma)$ ($\tilde{h}\in\tilde{G}$), the right-invariant 1-form $\rho(\sigma)$ is transformed as
\begin{align}
 \rho(\sigma) \to \rho(\sigma) + l\,\rmd\tilde{h}\,\tilde{h}^{-1}\,l^{-1}\,.
\end{align}
The gauge fields $\cC$ and $\cA$ and the intrinsic metric $\gamma$ are supposed to transform as
\begin{align}
 \cC(\sigma) \to \tilde{h}^{-1}\,\cC(\sigma)\,\tilde{h} - \tilde{h}^{-1}\,\rmd\tilde{h}\,,\qquad
 \cA(\sigma) \to \cA(\sigma)\,,\qquad
 \gamma \to \gamma\,.
\end{align}
We then find that $P(\sigma)$ and $\bm{\cP}(\sigma)$ are invariant under the right action. 
We can also find that the combination
\begin{align}
 \int_{\Sigma_2} \langle \rho \overset{\wedge}{,}\, l\,\cC\,l^{-1}\rangle 
 - \frac{1}{3!}\int_{\cB} \langle \rho\overset{\wedge}{,}\, [\rho,\, \rho]\rangle\,,
\end{align}
is invariant, and then the action \eqref{eq:DD-string-action} is gauge invariant. 

\item 
Let us denote the Lie group associated with $\mathfrak{f}$ as $F$. 
Under the left multiplication by an element of $F$, $l(\sigma)\to f(\sigma)\,l(\sigma)$ ($f\in F$), we find that the right-invariant 1-form transforms as 
\begin{align}
 \rho \to f\,\rho\,f^{-1} +\rmd f\,f^{-1}\,. 
\end{align}
The gauge fields are supposed to transform as
\begin{align}
 \cC(\sigma) \to \cC(\sigma)\,,\qquad
 \cA(\sigma) \to f(\sigma)\,\cA(\sigma)\,f^{-1}(\sigma) - \rmd f(\sigma)\,f^{-1}(\sigma) \,.
\end{align}
Then we can easily see that the combination
\begin{align}
 \int_{\Sigma_2} \bigl[\langle \rho \overset{\wedge}{,}\, l\,\cC\,l^{-1}\rangle - \langle P \overset{\wedge}{,}\, \cA\rangle\bigr]
 - \frac{1}{3!}\int_{\cB} \langle \rho\overset{\wedge}{,}\, [\rho,\, \rho]\rangle\,,
\end{align}
is invariant under the transformation. 
The first term of the action is slightly non-trivial. 
We find that $\bm{\cP}$ transforms as
\begin{align}
 \bm{\cP} \to f\,\bm{\cP}\,f^{-1} \quad \text{or}\quad 
 \bm{\cP}^A = (\text{Ad}_f)_B{}^A\,\bm{\cP}^B\,,
\end{align}
and then the action is invariant if the matrix $\hat{\cH}_{AB}\equiv \langle T_A,\,\hat{\cH}(T_B)\rangle$ satisfies
\begin{align}
 \hat{\cH}_{AB} = (\text{Ad}_f)_A{}^C\,(\text{Ad}_f)_B{}^D\,\hat{\cH}_{CD}\,,
\label{eqref:Killing-cond-finite}
\end{align}
for arbitrary $f\in F$ and the intrinsic metric $\gamma$ is invariant. 
For an infinitesimal gauge transformation $f(\sigma)=1+\epsilon^{\gi}(\sigma)\,T_{\gi}$\,, the condition \eqref{eqref:Killing-cond-finite} reads
\begin{align}
 \KK_{\gi}{}^C\,\bigl(F_{CA}{}^D\,\hat{\cH}_{DB} + F_{CB}{}^D\,\hat{\cH}_{AD}\bigr) = 0\,.
\label{eqref:Killing-cond}
\end{align}
\end{enumerate}

\subsection{Gauged string action in the second approach}

To consider the second approach, we shall begin by rewriting the action \eqref{eq:DD-string-action}. 
We then explain how to realize the gauge invariance in this approach. 

\subsubsection{Rewriting the action}

Assuming that an arbitrary element $l\in\cD$ can be factorized as $l=g\,\tilde{g}$ $(g\in G,\, \tilde{g}\in \tilde{G})$, we decompose the right-invariant 1-form field $\rho=\rmd l\,l^{-1}$ as
\begin{align}
 \rho \equiv \rho^A\,T_A = \rmd l \,l^{-1} = r + g\,\tilde{r}\,g^{-1} \,,
\end{align}
where we have defined
\begin{align}
 r\equiv r^a\,T_a \equiv \rmd g\,g^{-1}\,,\qquad
 \tilde{r}\equiv \tilde{r}_a\,T^a \equiv \rmd \tilde{g}\,\tilde{g}^{-1}\,.
\end{align}
We denote the adjoint action as $g\,T_A\,g^{-1} \equiv (\text{Ad}_g)_A{}^B\,T_B$\,, and then the matrix $\text{Ad}_{g}$ generally takes the form
\begin{align}
 (\text{Ad}_g)_A{}^B = \begin{pmatrix} (a^{-1})_a{}^b & 0 \\ a_c{}^a\,\pi^{cb} & a_b{}^a \end{pmatrix},
\label{eq:Ad-g}
\end{align}
where $a_b{}^a$ is invertible and $\pi^{ab}=-\pi^{ba}$\,. 
Using this parameterization, we can express the right-invariant 1-form and the gauge field $l\,\cC\,l^{-1}$ as
\begin{align}
 \rho^A = \begin{pmatrix} r^a - \pi^{ac}\,a_c{}^b\,\tilde{r}_b \\ a_a{}^b\,\tilde{r}_b \end{pmatrix},\qquad
 (l\,\cC\,l^{-1})^A = \begin{pmatrix} - \pi^{ab}\,\cC_b \\ \cC_a \end{pmatrix}.
\end{align}
We introduce the local coordinates on $G$ and $\tilde{G}$ as $g=\Exp{\delta_m^a x^m T_a}$ and $\tilde{g}=\Exp{\delta^m_a\tilde{x}_m T^a}$, respectively, and then we have
\begin{align}
 r^a=r^a_m(x)\,\rmd x^m\,,\qquad \tilde{r}_a = \tilde{r}_a^m(\tilde{x})\,\rmd \tilde{x}_m\,. 
\end{align}
We also introduce a matrix
\begin{align}
 E_M{}^A(x) \equiv \begin{pmatrix} r^a_m & 0 \\ -\pi^{ab}\,e_b^m & e_a^m \end{pmatrix},
\end{align}
where $e_a^m$ is the inverse of $r^a_m$.
By using this matrix, $\rho^A$ and $(l\,\cC\,l^{-1})^A$ can be rewritten as
\begin{align}
 \rho^A = E_M{}^A(x) \begin{pmatrix} \rmd x^m \\ \ell_m^a\,\tilde{r}_a \end{pmatrix},\qquad
 \cC^A = E_M{}^A(x) \begin{pmatrix} 0 \\ \ell_m^a(x)\,\cC_a \end{pmatrix},
\end{align}
where $\ell\equiv \ell_m^a\,T_a\,\rmd x^m\equiv g^{-1}\,\rmd g$ and we have used $e_a^m\,\ell_m^b=a_a{}^b$. 
Then, $P \equiv P^A\,T_A = \rho + l \,\cC \,l^{-1}$, which is defined in Eq.~\eqref{eq:defs}, can be expressed as
\begin{align}
 P^A(\sigma) = E_M{}^A\bigl(x(\sigma)\bigr)\,P^M(\sigma)\,,\qquad
 P^M(\sigma) \equiv \begin{pmatrix} \rmd x^m(\sigma) \\ P_m(\sigma) \end{pmatrix},
\end{align}
where $P_m \equiv \ell^a_m\,(C_a + \tilde{r}_a)$\,. 
We denote the inverse matrix of $E_M{}^A$ as $E_A{}^M$, and $P^M$ can be expressed as $P^M=E_A{}^M\,P^A$\,. 
Making a similar definition for $\bm{\cP}\equiv \bm{\cP}^A\,T_A$, we have
\begin{align}
 \bm{\cP}^M(\sigma) \equiv P^M(\sigma) + \cA^M(\sigma)\,,\qquad
 \cA^M(\sigma) \equiv \KK_{\gi}{}^M(x(\sigma))\,\cA^{\gi}(\sigma) \,, 
\end{align}
where $\KK_{\gi}{}^M(x) \equiv \KK_{\gi}{}^A\,E_A{}^M(x)$ are the generalized Killing vector fields that generate the gauge algebra $\mathfrak{f}$\,. 
Using the parameterization $(\KK_{\gi}{}^M)=(\kk_{\gi}^m,\,\tilde{\kk}_{\gi m})$, we can express $\bm{\cP}^M$ as
\begin{align}
 \bm{\cP}^M = \begin{pmatrix} \cD x^m \\ \bm{\cP}_m \end{pmatrix},\qquad
 \cD x^m \equiv \rmd x^m + \kk_{\gi}^m\,\cA^{\gi}\,,\qquad 
 \bm{\cP}_m\equiv P_m + \tilde{\kk}_{\gi m}\,\cA^{\gi}\,.
\end{align}
Further using the identity,
\begin{align}
 \int_{\Sigma_2} \langle \rho \overset{\wedge}{,}\, l\,\cC\,l^{-1}\rangle 
 - \frac{1}{3!}\int_{\cB} \langle \rho\overset{\wedge}{,}\, [\rho,\, \rho]\rangle
 = \int_{\Sigma_2} \rmd x^m\wedge P_m \,,
\end{align}
we can rewrite the gauged action \eqref{eq:DD-string-action} as
\begin{align}
 S = \frac{1}{4\pi\alpha'}\int_{\Sigma_2} \Bigl[\frac{1}{2}\,\cH_{MN}\bigl(x^m(\sigma)\bigr)\,\bm{\cP}^M \wedge *_\gamma \bm{\cP}^N - \eta_{MN}\, P^M \wedge \cA^N + \rmd x^m\wedge P_m\Bigr] \,,
\label{eq:DSM-coordinate}
\end{align}
where $\cH_{MN}\equiv E_M{}^A\,E_N{}^B\,\hat{\cH}_{AB}$ and $\eta_{MN}\equiv \Bigl(\begin{smallmatrix} 0 & \delta_m^n \\ \delta_m^n & 0 \end{smallmatrix}\Bigr)$. 

\subsubsection{Gauge symmetry}

Under the right multiplication $l(\sigma)\to l(\sigma)\,\tilde{h}(\sigma)$ $(\tilde{h}\in\tilde{G})$, $g=\Exp{\delta_m^a x^m T_a}$ and $\tilde{g}=\Exp{\delta^m_a\tilde{x}_m T^a}$ appearing in the decomposition $l=g\,\tilde{g}$ are transformed as
\begin{align}
 g'=g\quad \bigl(\Leftrightarrow\ \Exp{\delta_m^a x'^m T_a} = \Exp{\delta_m^a x^m T_a}\bigr),\qquad
 \tilde{g}'=\tilde{g}\,\tilde{h}\quad \bigl(\Leftrightarrow\ \Exp{\delta^m_a\tilde{x}'_m T^a} = \Exp{\delta^m_a\tilde{x}_m T^a}\tilde{h}\bigr),.
\end{align}
Namely, $x^m(\sigma)$ are invariant but $\tilde{x}_m(\sigma)$ are transformed. 
By recalling that $P(\sigma)=\rho + l \,\cC \,l^{-1}$ is invariant under the right multiplication, it is clear that the variation of $\tilde{x}_m$ (or $\tilde{r}_a$) and that of $\cC$ (or $C_a$) are canceled out, and their combination $P_m= \ell^a_m\,(C_a + \tilde{r}_a)$ is invariant. 
The gauge field $\cA^{\gi}$ is also invariant and then each term of the action \eqref{eq:DSM-coordinate} is manifestly invariant. 

The gauge symmetry associated with the left multiplication by $F$ is more non-trivial. 
To explain the details, let us explain some details on the left multiplication. 
As it has been noted in \cite{2112.14766}, an infinitesimal left multiplication by a constant element $(1+\epsilon^{\gi}\,T_{\gi})\in F$ corresponds to the generalized diffeomorphism along the generalized vector field $\epsilon^{\gi}\,\KK_{\gi}{}^M$. 
Under the condition \eqref{eqref:Killing-cond}, this generalized vector field satisfies the generalized Killing equations $\epsilon^{\gi}\gLie_{\KK_{\gi}}\cH_{MN} = 0$.
The generalized metric $\cH_{MN}$ contains the metric and the $B$-field as
\begin{align}
 \cH_{MN} = \begin{pmatrix} g_{mn}-B_{mp}\,g^{pq}\,B_{qn} & B_{mp}\,g^{pn} \\ -g^{mp}\,B_{pn} & g^{mn} 
\end{pmatrix},
\end{align}
and in terms of these fields, the generalized Killing equations read
\begin{align}
 \Lie_{\kk_{\gi}}g_{mn}=0\,,\qquad 
 \Lie_{\kk_{\gi}}B_2 + \rmd \tilde{\kk}_{\gi}^{(1)} =0\,,
\end{align}
where $\tilde{\kk}_{\gi}^{(1)}\equiv \tilde{\kk}_{\gi m}\,\rmd x^m$. 
The generalized Killing vector fields satisfy the algebra \cite{2112.14766}
\begin{align}
 \gLie_{\KK_{\gi}} \KK_{\gj}{}^I = - f_{\gi\gj}{}^{\gk}\,\KK_{\gk}{}^I\,,
\end{align}
and if we define $\hat{\kk}_{\gi m}$ through the relation,
\begin{align}
 \tilde{\kk}_{\gi m} = \hat{\kk}_{\gi m} + B_{mn}\,\kk_{\gi}^n\,.
\end{align}
the algebra can be expressed as
\begin{align}
 \Lie_{\kk_{\gi}}\kk_{\gj}^m = - f_{\gi\gj}{}^{\gk}\,\kk_{\gk}^m\,,\qquad 
 \Lie_{\kk_{\gi}}\hat{\kk}_{\gj}^{(1)} = - f_{\gi\gj}{}^{\gk}\,\hat{\kk}_{\gk}^{(1)}\,.
\end{align}

Now let us consider the gauge variation of the action \eqref{eq:DSM-coordinate}. 
Here we allow the parameter $\epsilon^{\gi}$ to depend on the worldsheet coordinates, and consider an infinitesimal diffeomorphism
\begin{align}
 \delta x^m(\sigma) = \epsilon^{\gi}(\sigma)\,\kk_{\gi}^m(x(\sigma))\,.
\label{eq:D-gauge1}
\end{align}
We also transform the gauge field $\cA$ as
\begin{align}
 \delta \cA^{\gi} = f_{\gj\gk}{}^{\gi}\,\epsilon^{\gj}\,\cA^{\gk} - \rmd \epsilon^{\gi}\,,
\label{eq:D-gauge2}
\end{align}
which is the same as the transformation rule in the first approach. 
The non-trivial point is the transformation rule of $P_m(\sigma)$. 
Here we redefine this as $\hat{P}_m(\sigma)$ through the relation
\begin{align}
 P_m = \hat{P}_m + B_{mn}\,\rmd x^n \,,
\end{align}
and treat $\hat{P}_m(\sigma)$ as the fundamental variable. 
We then assume the transformation rule
\begin{align}
 \delta \hat{P}_m = -\epsilon^{\gi}\,\partial_m k_{\gi}^n\,\hat{P}_n + \rmd \epsilon^{\gi}\,\hat{\kk}_{\gi m} \,.
\label{eq:D-gauge3}
\end{align}
Combining Eqs.~\eqref{eq:D-gauge1}--\eqref{eq:D-gauge3}, we can compute the variation of each term of the action. 
Any functions $f(x)$ of $x^m$ are transformed as $\delta f=\epsilon^{\gi}(\sigma)\,\kk_{\gi}^p\,\partial_p f(x)$\,, and then for example, we find
\begin{align}
\begin{split}
 \delta P_m &= \delta \hat{P}_m + \epsilon^{\gi}\,\kk_{\gi}^p\,\partial_p B_{mn}\,\rmd x^n + \rmd \epsilon^{\gi}\, B_{mn}\,\kk_{\gi}^n + B_{mp}\,\epsilon^{\gi}\,\partial_n \kk_{\gi}^p\,\rmd x^n
\\
 &= -\epsilon^{\gi}\,\partial_m k_{\gi}^n\,P_n
 + \rmd \epsilon^{\gi}\,\tilde{\kk}_{\gi m} 
 - 2\,\epsilon^{\gi}\, \partial_{[m}\,\tilde{k}_{|\gi|n]}\, \rmd x^n \,,
\\
 \delta \bm{\cP}^m &= \epsilon^{\gi}\,\partial_n\kk_{\gi}^m\,\bm{\cP}^n\,, \qquad
 \delta \bm{\cP}_m = -\epsilon^{\gi}\,\partial_m k_{\gi}^n\,\bm{\cP}_n
 - 2\,\epsilon^{\gi}\, \partial_{[m}\,\tilde{k}_{|\gi|n]}\, \bm{\cP}^n \,.
\end{split}
\end{align}
Then, the variation of the first term of the action is invariant
\begin{align}
 \delta\bigl(\cH_{MN}\,\bm{\cP}^M \wedge *_\gamma \bm{\cP}^N\bigr)=\epsilon^{\gi}\,(\gLie_{\KK_{\gi}}\cH)_{MN}\,\bm{\cP}^M \wedge *_\gamma \bm{\cP}^N = 0\,.
\end{align}
We can also find
\begin{align}
 \delta(-\eta_{MN}\, P^M \wedge \cA^N+\rmd x^m\wedge P_m\bigr) 
 = \rmd \bigl(-2\,\epsilon^{\gi}\, \tilde{\kk}_{\gi}^{(1)} \bigr)\,.
\end{align}
Then the action \eqref{eq:DSM-coordinate} is invariant under the gauge transformation up to the total derivative that corresponds to the $B$-field gauge transformation. 

To consider the case where the worldsheet has boundaries, we introduce an additional 1-form gauge field $A_1(\sigma)$ and add a total-derivative term $-\frac{1}{2\pi\alpha'}\int_{\Sigma_2} \rmd A_1$ to the action.
Then supposing that $A_1$ transforms as
\begin{align}
 \delta A_1 = -\epsilon^{\gi}\,\tilde{\kk}_{\gi}^{(1)} \,,
\end{align}
we can show that the resulting action is exactly invariant under the gauge transformation. 
If we denote the field strength of the gauge field as $F_2\equiv \rmd A_1$ and define a matrix $\omega^{(F)}$ as
\begin{align}
 \omega^{(F)}_{MN} \equiv \begin{pmatrix} 2\,F_{mn} & -\delta_m^n \\ \delta^m_n & 0 \end{pmatrix},
\end{align}
the modified action can be expressed as
\begin{align}
 S = \frac{1}{8\pi\alpha'}\int_{\Sigma_2} \bigl(\cH_{MN}\,\bm{\cP}^M\wedge *_\gamma\bm{\cP}^N 
 - \omega^{(F)}_{MN}\, P^M\wedge P^N 
 + 2\,\eta_{MN}\, \cA^M\wedge P^N\bigr) \,.
\label{eq:D-gBSM}
\end{align}
If we truncate the gauge field $\cA^{\gi}$, this is precisely the action of the Born sigma model
\begin{align}
 S = \frac{1}{8\pi\alpha'}\int_{\Sigma_2} \bigl(\cH_{MN}\, P^M\wedge *_\gamma P^N 
 - \omega^F_{MN}\, P^M\wedge P^N \bigr) \,,
\label{eq:D-BSM}
\end{align}
which has been studied in \cite{1910.09997,2004.09486}.
In this sense, the action \eqref{eq:D-gBSM} can be regarded as a gauged Born sigma model. 
In the following sections, we study extensions of the gauged Born sigma model for various branes in M-theory and type IIB theory. 

\section{Ungauged brane actions in M-theory/type IIB theory}
\label{sec:review-our-action}

The Born sigma model \eqref{eq:D-BSM} can be naturally generalized to various branes in M-theory and type IIB theory \cite{2004.09486}. 
In this section, we shortly review such generalizations. 
In the following, we shall use the multiple-index notation, which is reviewed in Appendix \ref{app:notation}.

The generalized Born sigma model for a $p$-brane takes the form \cite{2004.09486}
\begin{align}
 S = \frac{1}{2\,(p+1)} \int_{\Sigma_{p+1}} \Exp{\lambda}\cM_{IJ}(x) \,P^I\wedge *_{\gamma} P^J + \int_{\Sigma_{p+1}} \Omega_{\text{brane}} \,,
\end{align}
where
\begin{align}
 \Omega_{\text{brane}} \equiv -\frac{1}{2\,(p+1)}\,\omega^{(F)}_{IJ;\cK}\, P^I\wedge P^J\wedge q^\cK_{(\text{brane})}\,.
\end{align}
Let us explain the definition of each symbol one by one. 
\begin{enumerate}
\item 
The scalar field $\lambda(\sigma)$ is an auxiliary field that determines the tension of the brane when we consider a $(p,q)$-string in type IIB theory. 
When the dimensions of the worldvolume are different from two, $\lambda$ can be absorbed to the intrinsic metric $\gamma$ and can be ignored. 

\item 
The matrix $\cM_{IJ}(x)\in E_{\nM(\nM)}$ is the generalized metric (with unit determinant) in the $E_{\nM(\nM)}$ ExFT. 
This contains an overall determinant factor, but this factor can be absorbed into the auxiliary field $\lambda(\sigma)$ and can be ignored.
When we consider a brane in M-theory (type IIB theory), we assume that the generalized metric depends on $\nM$ ($\nM-1$) coordinates $x^i$ ($x^m$) ($i=1,\dotsc,\nM$ and $m=1,\dotsc,\nM-1$).
For later convenience, we denote the coordinates as $x^{\sfi}$ ($\sfi=1,\dotsc,D$ with $D=\nM$ in M-theory while $D=\nM-1$ in type IIB theory).
In both theories, the generalized metric can be decomposed as
\begin{align}
 \cM_{IJ} = (L^{\rmT}\,\hat{G}\,L)_{IJ}\,,
\label{eq:untwisted-metric}
\end{align}
where $\hat{G}_{IJ}$ is a block-diagonal matrix and a lower-triangular matrix $L^I{}_J$ (see Appendix \ref{app:Exceptional} for the details). 
The matrix $\hat{G}_{IJ}$ contains the metric and scalar fields in each theory while $L^I{}_J$ contains $p$-form (or more generally, mixed-symmetry) gauge potentials. 

We call the matrix $L^I{}_J$ a twist matrix, and accordingly, we call $\hat{G}_{IJ}$ the untwisted metric. 
We consider that an arbitrary generalized tensor field is twisted by the same twist matrix, and for example, for a given generalized vector field $V^I(x)$, we define the untwisted vector field $\hat{V}^I(x)$ as
\begin{align}
 V^I(x) = (L^{-1})^I{}_J(x)\,\hat{V}^J(x)\,, \qquad 
 \hat{V}^I(x) = L^I{}_J(x)\,\hat{V}^J(x)\,.
\end{align}
We note that the untwisted fields are very useful in our discussion. 

\item 
The 1-form fields $P^I$ take the following form in each theory:
\begin{align}
 P^I = \begin{pmatrix} \rmd x^i \\ P_{i_2} \\ P_{i_5} \\ \vdots\end{pmatrix}\quad \text{(M-theory)} ,\qquad 
 P^I = \begin{pmatrix} \rmd x^m \\ P^{\Ba}_m \\ P_{m_3} \\ P^{\Ba}_{m_5} \\ \vdots\end{pmatrix}\quad \text{(Type IIB theory)} .
\end{align}
The scalar fields $x^{\sfi}$ correspond to the embedding function and the other 1-form fields $P^{\cdots}_{\cdots}$ are auxiliary fields. 

\item 
The matrix $\omega^{(F)}_{IJ;\cK}$ is defined as
\begin{align}
 \omega^{(F)}_{IJ;\cK} \equiv (\cL^{\rmT})_I{}^K\,\omega^{(0)}_{KL;\cL}\,\cL^L{}_J\,\cL^{\cL}{}_{\cK}\,,
\label{eq:omega-F}
\end{align}
by using some constant matrices $\omega^{(0)}_{IJ;\cK}$\,. 
In M-theory, some of them (which are associated with M2 and M5) take the form
\begin{align}
 &\omega^{(0)}_k 
 = \begin{pmatrix}
 0 & - \delta_{ik}^{j_2} & 0 \\
 \delta_{jk}^{i_2} & 0 & 0 & \cdots \\
 0 & 0 & 0 \\ &\vdots&& \ddots 
 \end{pmatrix} , \qquad
 \omega^{(0)}_{k_4} 
 = \begin{pmatrix}
 0 & 0 & - \delta_{ik_4}^{j_5} \\
 0 & - \delta_{k_4}^{i_2j_2} & 0 & \cdots \\
 \delta_{jk_4}^{i_5} & 0 & 0 \\ &\vdots&& \ddots 
 \end{pmatrix} , \qquad \cdots\,.
\end{align}
In type IIB theory, some of them (which are associated with strings, D3, 5-branes) are
\begin{align}
 \omega^{(0) \Bc}&\equiv {\footnotesize\begin{pmatrix} 0 & -\delta_{\Bb}^{\Bc}\,\delta_{m}^{n} & 0 & 0 \\
 \delta_{\Ba}^{\Bc}\,\delta^{m}_{n} & 0 & 0 & 0 & \cdots \\
 0 & 0 & 0 & 0 \\
 0 & 0 & 0 & 0 \\ &\vdots &&& \ddots 
 \end{pmatrix} }, \quad 
 \omega^{(0)}_{p_2} \equiv {\footnotesize \begin{pmatrix}
 0 & 0 & - \delta_{mp_2}^{n_3} & 0 \\
 0 & - \epsilon_{\Ba\Bb}\,\delta_{p_2}^{nm} & 0 & 0 & \cdots \\
 \delta_{np_2}^{m_3} & 0 & 0 & 0 \\
 0 & 0 & 0 & 0 \\ &\vdots&&& \ddots 
 \end{pmatrix}},
\\
 \omega^{(0) \Bc}_{p_4} 
 &\equiv {\footnotesize {\arraycolsep=0mm \begin{pmatrix}
 0 & 0 & 0 & - \delta_{\Bb}^{\Bc}\,\delta_{p_4m}^{n_5} \\
 0 & 0 & - \delta_{\Ba}^{\Bc}\,\delta_{p_4}^{mn_3} & 0 & \cdots \\
 0 & - \delta_{\Bb}^{\Bc}\,\delta_{p_4}^{nm_3} & 0 & 0 \\
 \delta_{\Ba}^{\Bc}\,\delta_{p_4n}^{m_5} & 0 & 0 & 0 \\ &\vdots&&& \ddots 
 \end{pmatrix}}} ,\qquad \cdots.
\end{align}
The matrix $\cL^I{}_J\in E_{\nM(\nM)}$ contains the field strengths of the worldvolume gauge fields $F_{p+1}=\rmd A_{p}$ that extend $F_2$ appearing in the string action \eqref{eq:D-gBSM}, and $\cL^{\cI}{}_{\cJ}$ is the same matrix in a different representation of $E_{\nM(\nM)}$ (see Appendix \ref{app:Exceptional} for the details). 

\item 
The brane charges $q^\cI_{(\text{brane})}$ are defined as
\begin{align}
 q^\cI_{(\text{brane})} = (\cL^{-1})_{\cJ}{}^{\cI}\,\hat{q}^{\cJ}_{(\text{brane})}\,,
\end{align}
where $\hat{q}^{\cI}_{(\text{brane})}$ for M-theory branes are given by
\begin{align}
 \hat{q}^\cI_{(\text{M2})} \equiv \frac{\mu_2}{2}\begin{pmatrix} \rmd x^i \\ 0 \\ 0 \\ \vdots \end{pmatrix},\qquad
 \hat{q}^\cI_{(\text{M5})} \equiv \frac{\mu_5}{5}\begin{pmatrix} 0 \\ \rmd x^{i_4} \\ 0 \\ \vdots \end{pmatrix},\qquad\cdots,
\end{align}
and those for type IIB branes are given by
\begin{align}
 \hat{q}^\cI_{(p,q)\text{-1}} \equiv \mu_1{\footnotesize\begin{pmatrix} q_{\Ba} \\ 0 \\ 0 \\ 0 \\ \vdots \end{pmatrix}},\qquad
 \hat{q}^\cI_{(\text{D3})} \equiv \frac{\mu_3}{3}{\footnotesize\begin{pmatrix} 0 \\ \rmd x^{m_2} \\ 0 \\ 0 \\ \vdots \end{pmatrix}},
\qquad
 \hat{q}^\cI_{(p,q)\text{-5}} \equiv \frac{\mu_5}{5} {\footnotesize\begin{pmatrix} 0 \\ 0 \\ q_{\Ba}\,\rmd x^{m_4} \\ 0 \\ \vdots \end{pmatrix}},\quad\cdots.
\end{align}
Here, $\mu_p$ is a constant that corresponds to the tension (or charge) of the brane. 
The constants $q_{\Ba}$ ($\Ba=\bm{1},\bm{2}$) transform as an $\SL(2)$ $S$-duality doublet, and $q_{\Ba}=(1,0)$ corresponds to the F-string or the NS5-brane while $q_{\Ba}=(0,1)$ corresponds to the D-string or the D5-brane. 
\end{enumerate}

To find the explicit form of $\Omega_{\text{brane}}$, it is useful to define
\begin{align}
 \omega^{(0)}_{IJ} \equiv \omega^{(0)}_{IJ;\cK}\,\hat{q}^\cK_{(\text{brane})}\,,\qquad
 \omega^{(F)}_{IJ} \equiv (\cL^{\rmT} \omega^{(0)} \cL)_{IJ}\,,
\end{align}
and compute
\begin{align}
 \Omega_{\text{brane}} = -\frac{1}{2\,(p+1)}\, \omega^{(F)}_{IJ}\wedge P^I\wedge P^J \,.
\end{align}
Then, in M-theory, we find
\begin{align}
 \Omega_{(\text{M2})}
 &= - \tfrac{\mu_2}{3}\,P_{i_2}\wedge \rmd x^{i_2} - \mu_2\,F_3\,, 
\\
 \Omega_{(\text{M5})}
 &= - \tfrac{\mu_5}{6} \,\bigl(P_{i_5}\wedge \rmd x^{i_5} - P_{i_2}\wedge \rmd x^{i_2}\wedge F_3 \bigr)- \mu_5\,F_6 \,.
\end{align}
In type IIB theory, we find
\begin{align}
 \Omega_{(p,q)\text{-1}}
 &= - \tfrac{\mu_1}{2}\,q_{\Ba}\,P^{\Ba}_m\wedge \rmd x^m - \mu_1\,q_{\Ba}\, F_2^{\Ba} \,,
\\
 \Omega_{(\text{D3})} 
 &= - \tfrac{\mu_3}{4}\,\bigl(P_{m_3}\wedge \rmd x^{m_3} - \epsilon_{\Ba\Bb}\, P_m^{\Ba}\wedge \rmd x^m\wedge F_2^{\Bb}\bigr)
 - \mu_3\,F_4\,, 
\\
 \Omega_{(p,q)\text{-5}}
 &= -\tfrac{\mu_5}{6}\,q_{\Ba}\,\bigl(P^{\Ba}_{m_5}\wedge \rmd x^{m_5} - P_{m_3}\wedge \rmd x^{m_3}\wedge F_2^{\Ba} + P_{m}^{\Ba}\wedge \rmd x^{m} \wedge F_4 
\nn\\
 &\qquad\qquad\quad + \tfrac{1}{2}\, \epsilon_{\Bc\Bd}\,P^{\Bc}_m\wedge\rmd x^m\wedge F^{\Bd}_2\wedge F^{\Ba}_2 \bigr)
  - \mu_5\,q_{\Ba}\,\bigl(F_6^{\Ba} - \tfrac{2}{3}\,F_4 \wedge F_2^{\Ba} \bigr)\,.
\end{align}

Let us define the untwisted fields $\hat{P}^I$ as
\begin{align}
 \hat{P}^I \equiv L^I{}_J\,P^J\,.
\end{align}
In M-theory, we have
\begin{align}
\begin{split}
 P^i &= \hat{P}^i = \rmd x^i\,, \qquad
 P_{i_2} = \hat{P}_{i_2} - C_{i_2j}\,\rmd x^j\,,
\\
 P_{i_5} &= \hat{P}_{i_5} + C_{[i_3}\,\hat{P}_{i_2]} + \bigl(C_{i_5j}-\tfrac{1}{2}\,C_{[i_3}\,A_{i_2]j}\bigr)\,\rmd x^j\,,
\end{split}
\end{align}
where $C_{3}$ and $C_{6}$ are the standard 3- and 6-form potential in eleven-dimensional supergravity. 
In type IIB theory, we have
\begin{align}
\begin{split}
 P^m &= \hat{P}^{m} = \rmd x^m\,, \qquad
 P^{\Ba}_{m} = \hat{P}^{\Ba}_{m} + B^{\Ba}_{mn}\,\rmd x^n \,,
\\
 P_{m_3} &= \hat{P}_{m_3} - \epsilon_{\Bc\Bd}\,B^{\Bc}_{[m_2}\,\hat{P}^{\Bd}_{m_1]} + \bigl(B_{m_3n}+\tfrac{1}{2}\,\epsilon_{\Bc\Bd}\,B^{\Bc}_{[m_2}\,B^{\Bd}_{m_1]n})\,\rmd x^n\,,
\\
 P^{\Ba}_{m_5} &= \hat{P}^{\Ba}_{m_5} + B^{\Ba}_{[m_2}\,\hat{P}_{m_3]} - B_{[m_4}\,\hat{P}^{\Ba}_{m_1]}
 - \tfrac{1}{2}\,\epsilon_{\Bc\Bd}\,B^{\Ba}_{[m_2}\,B^{\Bc}_{m_2}\,\hat{P}^{\Bd}_{m_1]}
\nn\\
 &\quad + \bigl(B^{\Ba}_{m_5n} - B_{[m_4}\,B^{\Ba}_{m_1]n} - \tfrac{1}{3!}\,\epsilon_{\Bc\Bd}\,B^{\Ba}_{[m_2}\,B^{\Bc}_{m_2}\,B^{\Bd}_{m_1]n}\bigr)\,\rmd x^n\,,
\end{split}
\end{align}
where $B^{\Ba}_{2}$, $B_{4}$, and $B^{\Ba}_6$ are $S$-duality-covariant potential fields in type IIB supergravity. 

By using the untwisted fields, in M-theory, the kinetic term is simplified as
\begin{align}
 \cM_{IJ}\,P^I\wedge *_\gamma P^J = g_{ij}\,\hat{P}^i\wedge *_\gamma\hat{P}^j + g^{i_2j_2}\,\hat{P}_{i_2}\wedge *_\gamma\hat{P}_{j_2} + g^{i_5j_5}\,\hat{P}_{i_5}\wedge *_\gamma\hat{P}_{j_5}+\cdots\,.
\label{eq:kinetic-infiniteM}
\end{align}
In type IIB theory, we have
\begin{align}
\begin{split}
 \cM_{IJ}\,P^I\wedge *_\gamma P^J &= g_{mn}\,\hat{P}^m\wedge *_\gamma\hat{P}^n + m_{\Ba\Bb}\,g^{mn}\,\hat{P}^{\Ba}_{m}\wedge *_\gamma\hat{P}^{\Bb}_{n}
\nn\\
 &\quad + g^{m_3n_3}\,\hat{P}_{m_3}\wedge *_\gamma\hat{P}_{m_3} + m_{\Ba\Bb}\,g^{m_5n_5}\,\hat{P}^{\Ba}_{m_5}\wedge *_\gamma\hat{P}^{\Bb}_{n_5} +\cdots\,.
\end{split}
\label{eq:kinetic-infiniteB}
\end{align}
When $\nM\leq 8$, there are only finite terms, but when $\nM\geq 9$, there is an infinite number of terms. 

We can also rewrite $\Omega_{\text{brane}}$ as follows by using the untwisted fields:
\begin{align}
 \Omega_{(\text{M2})}
 &= -\tfrac{\mu_2}{3}\,\hat{P}_{i_2}\wedge \rmd x^{i_2} + \cL^{\text{WZ}}_{(\text{M2})}\,, 
\\
 \Omega_{(\text{M5})}
 &= - \tfrac{\mu_5}{6}\,\bigl[\hat{P}_{i_5}\wedge \rmd x^{i_5} + \hat{P}_{i_2}\wedge \rmd x^{i_2}\wedge (A_3-F_3)\bigr]
 + \cL^{\text{WZ}}_{(\text{M5})} \,,
\\
 \vdots&
\nn\\
 \Omega_{(p,q)\text{-1}}
 &= -\tfrac{\mu_1}{2}\,q_{\Ba}\,\hat{P}^{\Ba}_m\wedge \rmd x^m + \cL^{\text{WZ}}_{(p,q)\text{-1}}\,,
\\
 \Omega_{(\text{D3})} 
 &= - \tfrac{\mu_3}{4}\,\bigl[\hat{P}_{m_3}\wedge \rmd x^{m_3} 
  + \epsilon_{\Ba\Bb}\, \hat{P}_m^{\Ba}\wedge \rmd x^m\wedge (B_2^{\Bb}-F_2^{\Bb})\bigr] + \cL^{\text{WZ}}_{(\text{D3})}\,, 
\\
 \Omega_{(p,q)\text{-5}}
 &= -\tfrac{\mu_5}{6}\,q_{\Ba}\,\bigl[\hat{P}^{\Ba}_{m_5}\wedge \rmd x^{m_5} + \hat{P}_{m_3}\wedge \rmd x^{m_3}\wedge (B_2^{\Ba}-F_2^{\Ba}) - \hat{P}_{m}^{\Ba}\wedge \rmd x^{m} \wedge (B_4-F_4)
\nn\\
 &\qquad\qquad\quad + \tfrac{1}{2}\, \epsilon_{\Bc\Bd}\,\hat{P}^{\Bc}_m\wedge\rmd x^m\wedge (B^{\Bd}_2\wedge B^{\Ba}_2 -2\,B^{\Bd}_2\wedge F^{\Ba}_2 +F^{\Bd}_2\wedge F^{\Ba}_2) \bigr] + \cL^{\text{WZ}}_{(p,q)\text{-5}}\,,
\\
 \vdots&\,,
\nn
\end{align}
where $\cL^{\text{WZ}}_{\text{brane}}$ denotes the Wess--Zumino (WZ) term
\begin{align}
\begin{split}
 \cL^{\text{WZ}}_{(\text{M2})}
 &\equiv \mu_2\,(A_3 - F_3)\,, \qquad
 \cL^{\text{WZ}}_{(\text{M5})}
  \equiv \mu_5\,\bigl(A_6-\tfrac{1}{2}\,A_3\wedge F_3 - F_6 \bigr)\,,
\\
 \cL^{\text{WZ}}_{(p,q)\text{-1}}
 &\equiv \mu_1\,q_{\Ba}\,\bigl(B^{\Ba}_2 - F_2^{\Ba} \bigr)\,,\qquad
 \cL^{\text{WZ}}_{(\text{D3})} 
  \equiv \mu_3\,\bigl(B_4 -\tfrac{1}{2}\,\epsilon_{\Ba\Bb}\,B_2^{\Ba}\wedge F_2^{\Bb} - F_4\bigr)\,, 
\\
 \cL^{\text{WZ}}_{(p,q)\text{-5}}
 &\equiv \mu_5\,q_{\Ba}\,\bigl[B_6^{\Ba} - F_6^{\Ba} -(B_4-F_4)\wedge (\tfrac{1}{3}\,B_2^{\Ba}+\tfrac{2}{3}\, F_2^{\Ba}) +\tfrac{1}{6}\,\epsilon_{\Bc\Bd}\,B_2^{\Bc} \wedge F_2^{\Bd}\wedge F_2^{\Ba} \bigr] \,.
\end{split}
\label{eq:WZ}
\end{align}
Using these expressions, we can easily write down the brane action for each brane. 

In summary, the action can be expressed schematically as
\begin{align}
 S = \int_{\Sigma_{p+1}}\Bigl[\tfrac{1}{2\,(p+1)} \Exp{\lambda}\cM_{IJ}(x) \,P^I\wedge *_{\gamma} P^J - \tfrac{\mu_p}{p+1}\,\hat{P}_{\sfi_p}\wedge \rmd x^{\sfi_p} + \cdots \Bigr] + S^{\text{WZ}} \,.
\label{eq:action-rough}
\end{align}
The kinetic term $\cM_{IJ}(x) \,P^I\wedge *_{\gamma} P^J$ contains an infinite number of terms if $\nM$ is greater than 8 (recall Eqs.~\eqref{eq:kinetic-infiniteM} and \eqref{eq:kinetic-infiniteB}). 
However, when we consider a $p$-brane, the auxiliary fields $\hat{P}^{\cdots}_{\cdots}$ with the number of the curved indices greater than $p$ appear only from the kinetic term. 
Then, they always have simple quadratic forms, such as $g^{\sfi_q\sfj_q}\,\hat{P}_{\sfi_q}\wedge*_\gamma\hat{P}_{\sfj_q}$, and can be eliminated under the equations of motion for the auxiliary fields. 
We thus argue that $\nM$ can be extended to $\nM=11$ when we consider standard branes, such as M2-brane or M5-brane as well as the standard type IIB branes. 
In ExFT, we need to decompose the elven- or ten-dimensional spacetime into the external space and the internal space, and our coordinates $x^{\sfi}$ correspond to those of the internal space. 
The reason why we do not consider the external space is because we expect that $\nM$ can be extended to $\nM=11$, where the external space vanishes. 

We note that, after eliminating all of the auxiliary fields $\lambda$, $\gamma$, and $P^{\cdots}_{\cdots}$, the first term of the action \eqref{eq:action-rough}, $\int_{\Sigma_{p+1}}\bigl[\cdots \bigr]$, reproduces the kinetic term of the Nambu--Goto type.
Then $S^{\text{WZ}}$ exactly corresponds to the WZ term. 
Indeed, for a membrane, the M5-brane, and the $(p,q)$-string, the WZ terms given in Eq.~\eqref{eq:WZ} are the standard ones. 
However, for the D3-brane, our WZ term looks different from the standard one. 
Our WZ term rather corresponds to that of the $S$-duality-covariant formulation of D3-brane worldvolume theory (see, for example, Eq.~(6.20) of \cite{1712.06425}). 
The WZ term of the $(p,q)$-5-brane also seems different from the standard one but we expect that our theory is equivalent to the standard one. 
In our approach, for each potential $C_p$ in supergravity, there is the corresponding field strength $F_p$ of the worldvolume gauge fields. 
This is because both of the potentials and the field strengths are associated with certain generators of the exceptional algebra $\mathfrak{e}_{\nM(\nM)}$ (see Appendix \ref{app:Exceptional}). 
In section \ref{sec:gauged-brane}, we see that this $U$-duality covariance helps us to construct the gauged actions. 

\section{Gauged brane actions}
\label{sec:gauged-brane}

In this section, we construct gauged brane actions that describe the dynamics of branes on the exceptional dressing cosets. 

\subsection{Exceptional dressing cosets}

The brane sigma model reviewed in the previous section describes the dynamics of a certain brane in a curved background with the supergravity fields $\cM_{IJ}(x^{\sfi})$. 
When we discuss the generalized $U$-duality, we construct $\cM_{IJ}$ by using the inverse of the generalized frame fields $E_A{}^I(x^{\sfi})$ associated with EDA and a constant matrix $\hat{\cM}_{AB}\in E_{\nM(\nM)}$ as \cite{1911.06320,1911.07833}
\begin{align}
 \cM_{IJ}(x)\propto E_I{}^A(x)\,E_J{}^B(x)\,\hat{\cM}_{AB}\,.
\end{align}
Then the target space is a $D$-dimensional background $M$ that is a $U$-dual version of the PL $T$-dualizable background $M=\cD/\tilde{G}$\,. 
By the construction, supergravity fields depend only on the coordinates $x^{\sfi}$ on the Lie group $G$, and the target space $M$ can be identified with $G$. 
To construct the exceptional dressing cosets, we assume that this background $M$ admits $n$ generalized Killing vector fields $\KK_{\gi}{}^I\equiv \KK_{\gi}{}^A\,E_A{}^I$ ($\gi=1,\dotsc,n$, $\KK_{\gi}{}^A:\text{constants}$)
\begin{align}
 \gLie_{\KK_{\gi}}\cM_{IJ} = 0\,,
\end{align}
where $\KK_{\gi}$ are supposed to form a algebra $\mathfrak{f}$
\begin{align}
 \gLie_{\KK_{\gi}}\KK_{\gj}{}^I = -f_{\gi\gj}{}^{\gk}\,\KK_{\gk}{}^I\,.
\end{align}
We also assume that $\mathfrak{f}$ is isotropic (which ensures that $\mathfrak{f}$ is a Lie algebra), and define the dressing action as generalized diffeomorphisms along these isometry directions. 
An infinitesimal dressing action on $M$ is generated by $\epsilon^{\gi} \gLie_{\KK_{\gi}}$ ($\epsilon^{\gi}:\text{constants}$). 
Then, in the sigma model, we promote this global symmetry into the local gauge symmetry, and the dressing action is gauged. 
As the result of this gauging, the target space is reduced to the coset $F\backslash M\simeq F\backslash G$\,, which is the exceptional dressing coset. 

Note that if we define $T_{\gi}\equiv \KK_{\gi}{}^A\,T_A$\,, they form a Lie subalgebra $\mathfrak{f}$\,: $[T_{\gi},\,T_{\gj}]=f_{\gi\gj}{}^{\gk}\,T_{\gk}$\,.
In particular, if $\mathfrak{f}$ is a subalgebra of $\mathfrak{g}$\,, the dressing action is generated by the right-invariant vector fields $e_{\gi}^m$ on $M\simeq G$ and the exceptional dressing coset reduces to the usual coset $F\backslash G$. 

\subsection{Our conventions}

Before going into details, let us explain our conventions on supergravity fields. 
In ExFT, the generalized Lie derivative generates the usual Lie derivative and the gauge transformations of various potential fields. 
In the M-theory section, if we parameterize the diffeomorphism parameters as $(V^I)=(v^i,\,\tilde{v}_{i_2},\,\tilde{v}_{i_5},\,\cdots)$\,, the variation of supergravity fields are given by
\begin{align}
 \delta g_{ij} = \Lie_v g_{ij}\,,
\quad
 \delta C_3 = \Lie_v C_3 + \rmd \tilde{v}_{2} \,,
\quad
 \delta C_6 = \Lie_v C_6 + \rmd \tilde{v}_{5} + \tfrac{1}{2}\,C_3\wedge \rmd \tilde{v}_2 \,, \qquad \cdots \,.
\label{eq:M-gKE}
\end{align}
In the type IIB section, using a parameterization $(V^I)=(v^m,\,\tilde{v}^{\Ba}_m,\,\tilde{v}_{m_3},\,\tilde{v}^{\Ba}_{m_5},\,\cdots)$, we have
\begin{align}
\begin{split}
 \delta \mathsf{g}_{mn} &= \Lie_v \mathsf{g}_{mn}\,,\quad
 \delta m_{\Ba\Bb} = \Lie_v m_{\Ba\Bb}\,, \quad
 \delta B^{\Ba}_2 = \Lie_v B^{\Ba}_2 + \rmd \tilde{v}_1^{\Ba} \,,
\\
 \delta B_4
 &= \Lie_v B_4 + \rmd \tilde{v}_3 + \tfrac{1}{2}\,\epsilon_{\Bc\Bd}\,B^{\Bc}_2\wedge \rmd \tilde{v}_1^{\Bd}\,, 
\\
 \delta B^{\Ba}_6
  &= \Lie_v B^{\Ba}_6 + \rmd \tilde{v}^{\Ba}_5 +B_4 \wedge\rmd\tilde{v}^{\Ba}_1
 + \tfrac{1}{3!}\,\epsilon_{\Bc\Bd}\,B^{\Ba}_2\wedge B^{\Bc}_2\wedge\rmd\tilde{v}^{\Bd}_1 \,,
\end{split}
\end{align}
where $\mathsf{g}_{mn}$ is the Einstein-frame metric and $m_{\Ba\Bb}$ contains the dilaton and the Ramond--Ramond 0-form potential. 
Then, for example in M-theory, if we parameterize the generalized Killing vector fields as $(\KK_{\gi}{}^I)=(\kk_{\gi}^i,\,\tilde{\kk}_{\gi i_2},\,\tilde{\kk}_{\gi i_5},\,\cdots)$, the generalized Killing equations can be expressed as
\begin{align}
 \Lie_{\kk_{\gi}}g_{ij} =0\,,\quad 
 \Lie_{\kk_{\gi}}C_3 + \rmd \tilde{\kk}^{(2)}_{\gi} = 0 \,, \quad
 \Lie_{\kk_{\gi}}C_6 + \rmd \tilde{\kk}_{\gi}^{(5)} + \tfrac{1}{2}\,C_3\wedge \rmd \tilde{\kk}^{(2)}_{\gi}=0 \,, \qquad \cdots \,,
\end{align}
where $\hat{\kk}^{(p)}_{\gi}\equiv \hat{\kk}_{\gi i_p}\,\rmd x^{i_p}$. 

For convenience, we shall define the untwisted vector fields $\hat{\KK}_{\gi}{}^I$ as
\begin{align}
 \KK_{\gi}{}^I = (L^{-1})^I{}_J\, \hat{\KK}_{\gi}{}^J \,.
\end{align}
In M-theory/type IIB theory, they are parameterized as
\begin{align}
 \hat{\KK}_{\gi}{}^I \equiv \begin{pmatrix} \kk_{\gi}^i \\ \hat{\kk}_{\gi i_2} \\ \hat{\kk}_{\gi i_5} \\ \vdots \end{pmatrix},\qquad
 \hat{\KK}_{\gi}{}^I \equiv \begin{pmatrix} \kk_{\gi}^m \\ \hat{\kk}^{\Ba}_{\gi m} \\ \hat{\kk}_{\gi m_3} \\ \hat{\kk}^{\Ba}_{\gi m_5} \\ \vdots \end{pmatrix}.
\end{align}
In terms of these untwisted fields, the algebra $\gLie_{\KK_{\gi}}\KK_{\gj}{}^I=-f_{\gi\gj}{}^{\gk}\,\KK_{\gk}{}^I$ can be simplified as
\begin{align}
 \Lie_{\kk_{\gi}}\kk_{\gj}^{\sfi} = -f_{\gi\gj}{}^{\gk}\,\kk_{\gk}^{\sfi}\,,\qquad 
 \Lie_{\kk_{\gi}}\hat{\kk}^{(p)}_{\gj} = -f_{\gi\gj}{}^{\gk}\,\hat{\kk}_{\gk}^{(p)}\,,\qquad
 \Lie_{\kk_{\gi}}\hat{\kk}^{\Ba(p)}_{\gj} = -f_{\gi\gj}{}^{\gk}\,\hat{\kk}_{\gk}^{\Ba(p)}\,.
\end{align}
By using the explicit form of the matrix $L^{-1}\equiv (L^{-1})^I{}_J$ in M-theory
\begin{align}
 L^{-1} = \begin{pmatrix}
 \delta^i_j & 0 & 0 & 
\\
 - C_{i_2 j} & \delta_{i_2}^{j_2} & 0 & \cdots
\\
 C_{i_5 j} - \frac{1}{2!}\,C_{[i_3}\,C_{i_2] j} & \delta^{j_2k_3}_{i_5}\,C_{k_3} & \delta_{i_5}^{j_5} & \\
 \vdots &&&\ddots
 \end{pmatrix},
\end{align}
and in type IIB theory
\begin{align}
 L^{-1} = {\tiny\begin{pmatrix}
 \delta^m_n & 0 & 0 & 0 & 
\\[1mm]
 B^{\Ba}_{mn} & \delta^{\Ba}_{\Bb}\delta_m^n & 0 & 0 & \ldots 
\\[1mm]
 B_{m_3n}-\frac{1}{2}\epsilon_{\Bc\Bd}B^{\Bc}_{[m_2}B^{\Bd}_{m_1]n} & \epsilon_{\Bb\Bc}B^{\Bc}_{[m_2}\delta^{n}_{m_1]} & \delta_{m_3}^{n_3} & 0 & 
\\[1mm]
 B^{\Ba}_{m_5n} - B_{[m_4}B^{\Ba}_{m_1]n} - \frac{1}{3!}\epsilon_{\Bc\Bd}B^{\Ba}_{[m_2}B^{\Bc}_{m_2}B^{\Bd}_{m_1]n} & -\delta^{\Ba}_{\Bb}\delta^{n}_{[m_1}B_{m_4]} + \frac{1}{2}\epsilon_{\Bb\Bc}\delta^{n}_{[m_1}B^{\Ba}_{m_2}B^{\Bc}_{m_2]} & B^{\Ba}_{[m_2}\delta^{n_3}_{m_3]} & \delta^{\Ba}_{\Bb}\delta_{m_5}^{n_5} & \\
 \vdots &&&& \ddots
\end{pmatrix}},
\end{align}
we can parameterize the generalized Killing vector fields in M-theory as
\begin{align}
 \KK_{\gi}{}^I \equiv {\footnotesize\begin{pmatrix} \kk_{\gi}^i 
\\[1mm]
 \hat{\kk}_{\gi i_2} - C_{i_2j}\kk_{\gi}^j
\\[1mm]
 \hat{\kk}_{\gi i_5} + C_{[i_3} \hat{\kk}_{|\gi|i_2]} + (C_{i_5j}-\frac{1}{2}\,C_{[i_3}C_{i_2]j})\kk^j_{\gi}\\ \vdots 
\end{pmatrix}},
\end{align}
and in type IIB theory as
\begin{align}
 \KK_{\gi}{}^I \equiv {\tiny\begin{pmatrix} \kk_{\gi}^m 
\\[1mm]
 \hat{\kk}^{\Ba}_{\gi m} + B^{\Ba}_{mn}\kk_{\gi}^n 
\\[1mm]
 \hat{\kk}_{\gi m_3} - \epsilon_{\Bc\Bd}B^{\Bc}_{[m_2} \hat{\kk}^{\Bd}_{|\gi|m_1]} + (B_{m_3n}-\frac{1}{2}\epsilon_{\Bc\Bd}B^{\Bc}_{[m_2}B^{\Bd}_{m_1]n})\kk^n_{\gi} 
\\[1mm]
 \hat{\kk}^{\Ba}_{\gi m_5} + B^{\Ba}_{[m_2}\hat{\kk}_{|\gi|m_3]} - B_{[m_4}\hat{\kk}^{\Ba}_{|\gi|m_1]} - \frac{1}{2}\epsilon_{\Bc\Bd} B^{\Ba}_{[m_2}B^{\Bc}_{m_2}\hat{\kk}^{\Bd}_{|\gi|m_1]}
 + (B^{\Ba}_{m_5n} - B_{[m_4}B^{\Ba}_{m_1]n} - \frac{1}{3!}\epsilon_{\Bc\Bd}B^{\Ba}_{[m_2}B^{\Bc}_{m_2}B^{\Bd}_{m_1]n})\kk^n_{\gi} \\ \vdots \end{pmatrix}}.
\end{align}

The gauge algebra $\mathfrak{f}$ has been assumed to be isotropic, which can be expressed as
\begin{align}
 \KK_{\gi}{}^A\,\KK_{\gj}{}^B\,\eta_{AB;\cI} =0 \,,\qquad
 \cdots\,,
\end{align}
where $\eta_{AB;\cI}$ is the so-called $\eta$-symbol that defines the section condition in ExFT, and the ellipses become important only when we consider exotic branes. 
In M-theory, this condition is equivalent to
\begin{align}
 0&=\iota_{\gi}\hat{\kk}^{(2)}_{\gj} + \iota_{\gj}\hat{\kk}^{(2)}_{\gi} \,,
\label{eq:M2-SC}
\\
 0&=\iota_{\gi}\hat{\kk}^{(5)}_{\gj}+ \hat{\kk}^{(2)}_{\gi}\wedge\hat{\kk}^{(2)}_{\gj}+\iota_{\gj}\hat{\kk}^{(5)}_{\gi} \,,\quad \cdots\,,
\label{eq:M5-SC}
\end{align}
where $\iota_{\gi}\equiv \iota_{\kk_{\gi}}$\,. 
In type IIB theory, this is equivalent to
\begin{align}
 0&=\iota_{\gi}\hat{\kk}^{\Ba(1)}_{\gj} + \iota_{\gj}\hat{\kk}^{\Ba(1)}_{\gi}\,,
\label{eq:B1-SC}
\\
 0&=\iota_{\gi}\hat{\kk}^{(3)}_{\gj}+\iota_{\gj}\hat{\kk}^{(3)}_{\gi}
 -\epsilon_{\Ba\Bb}\,\hat{\kk}^{\Ba(1)}_{\gi}\wedge\hat{\kk}^{\Bb(1)}_{\gj} \,, 
\label{eq:B3-SC}
\\
 0&=\iota_{\gi}\hat{\kk}^{\Ba(5)}_{\gj}
 + \hat{\kk}^{\Ba(1)}_{\gi}\wedge\hat{\kk}^{(3)}_{\gj}
 + \hat{\kk}^{\Ba(1)}_{\gj}\wedge\hat{\kk}^{(3)}_{\gi}
 +\iota_{\gj}\hat{\kk}^{\Ba(5)}_{\gi} \,,\quad \cdots\,.
\label{eq:B5-SC}
\end{align}
We note that these isotropy conditions are very important when we construct the gauge-invariant WZ terms in our gauged brane actions.

\subsection{Gauge transformations}

To construct the gauged action, we here define the gauge transformation of various fields on the worldvolume and construct several gauge-invariant combinations. 

We suppose that the coordinates $x^{\sfi}$ and the gauge fields $\cA^{\gi}$ transform as
\begin{align}
 \delta x^{\sfi} = \epsilon^{\gi}\,\kk_{\gi}^{\sfi}\,,\qquad 
 \delta \cA^{\gi} = -\rmd \epsilon^{\gi} + f_{\gj\gk}{}^{\gi}\,\epsilon^{\gj}\,\cA^{\gk}\,. 
\end{align}
They are natural generalizations of Eqs.~\eqref{eq:D-gauge1} and \eqref{eq:D-gauge2}. 
If we define $Dx^{\sfi}\equiv\rmd x^{\sfi}+\cA^{\gi}\,\kk_{\gi}^{\sfi}$\,, we can easily find
\begin{align}
 \delta Dx^{\sfi} = \epsilon^{\gi}\,\partial_{\sfj}\kk_{\gi}^{\sfi}\,Dx^{\sfj} \,.
\end{align}
In addition, we need to assume the transformations of the auxiliary fields $\hat{P}^{\cdots}_{\cdots}$ similar to the one given in Eq.~\eqref{eq:D-gauge3}. 
In the M-theory case, we suppose
\begin{align}
 \delta \hat{P}_{i_2} &= -\epsilon^{\gi}\,\delta_{i_2}^{kl}\,\partial_{k}\kk_{\gi}^j\,\hat{P}_{jl} + \rmd \epsilon^{\gi}\,\hat{\kk}_{\gi i_2} \,,
\\
 \delta \hat{P}_{i_5} &= -\epsilon^{\gi}\,\delta_{i_5}^{kl_4}\,\partial_{k}\kk_{\gi}^j\,\hat{P}^{\Ba}_{jl_4} + \rmd \epsilon^{\gi}\,\hat{\kk}^{\Ba}_{\gi i_5}\,,\quad \cdots\,,
\end{align}
while in the type IIB case, we suppose
\begin{align}
 \delta \hat{P}^{\Ba}_m &= -\epsilon^{\gi}\,\partial_m\kk_{\gi}^n\,\hat{P}^{\Ba}_n + \rmd \epsilon^{\gi}\,\hat{\kk}^{\Ba}_{\gi m}\,,
\\
 \delta \hat{P}_{m_3} &= -\epsilon^{\gi}\,\delta_{m_3}^{pq_2}\,\partial_{p}\kk_{\gi}^n\,\hat{P}_{nq_2} + \rmd \epsilon^{\gi}\,\hat{\kk}_{\gi m_3} \,,
\\
 \delta \hat{P}^{\Ba}_{m_5} &= -\epsilon^{\gi}\,\delta_{m_5}^{pq_4}\,\partial_{p}\kk_{\gi}^n\,\hat{P}^{\Ba}_{nq_4} + \rmd \epsilon^{\gi}\,\hat{\kk}^{\Ba}_{\gi m_5}\,,\quad \cdots\,.
\end{align}
In both cases, these can be neatly summarized as (see Appendix \ref{app:Exceptional} for the definition of $\tilde{K}^{\sfi}{}_{\sfj}$)
\begin{align}
 \delta \hat{P}^I = - \epsilon^{\gi}\,\partial_{\sfi} \kk_{\gi}^{\sfj}\,(\tilde{K}^{\sfi}{}_{\sfj})^I{}_J\,\hat{P}^J + \rmd \epsilon^{\gi}\,\hat{\KK}_{\gi}{}^I\,. 
\end{align}
We assume that the scalar field $\lambda(\sigma)$ and the intrinsic metric on the worldvolume, which defines $*_\gamma$, are invariant under the gauge transformation. 

In the sigma model, we regard $\hat{\KK}_{\gi}{}^I(x(\sigma))$ as functions, which transform as
\begin{align}
 \delta \hat{\KK}_{\gi}{}^I(x) = \epsilon^{\gi}\,\kk_{\gi}^{\sfi}\,\partial_{\sfi} \hat{\KK}_{\gi}{}^I(x)\,.
\end{align}
Then, we find that
\begin{align}
 \hat{\bm{\cP}}^I \equiv \hat{P}^I + \hat{\KK}_{\gi}{}^I\,\cA^{\gi}\,,
\end{align}
transforms as
\begin{align}
 \delta \hat{\bm{\cP}}^I = - \epsilon^{\gi}\,\partial_{\sfi} \kk_{\gi}^{\sfj}\,(\tilde{K}^{\sfi}{}_{\sfj})^I{}_J\,\hat{\bm{\cP}}^J\,, 
\label{eq:delta-bP}
\end{align}
where we have used
\begin{align}
 \kk_{\gi}^{\sfi}\,\partial_{\sfi} \hat{\KK}_{\gj}{}^I 
+ \partial_{\sfi} \kk_{\gi}^{\sfj}\,(\tilde{K}^{\sfi}{}_{\sfj})^I{}_J\,\hat{\KK}_{\gj}{}^J 
 = - f_{\gi\gj}{}^{\gk}\,\hat{\KK}_{\gk}{}^I \,.
\end{align}

The generalized metric $\cM_{IJ}$ can be decomposed as \eqref{eq:untwisted-metric} and the untwisted metric $\hat{G}_{IJ}$ contains only the metric (or scalar fields) which are invariant under the gauge transformations of $p$-form potentials. 
Namely, the generalized Killing equations give
\begin{align}
 \kk_{\gi}^{\sfk}\,\partial_{\sfk} \hat{G}_{IJ}- \partial_{\sfk} \kk_{\gi}^{\sfl} \, \bigl[(\tilde{K}^{\sfk}{}_{\sfl})^{\rmT}\,\hat{G} + \hat{G}\,\tilde{K}^{\sfk}{}_{\sfl}\bigr]_{IJ} = 0\,.
\end{align}
Using this, the variation of the untwisted metric under the gauge transformation becomes
\begin{align}
 \delta \hat{G}_{IJ}= \epsilon^{\gi}\,\kk_{\gi}^{\sfk}\,\partial_{\sfk} \hat{G}_{IJ} = \partial_{\sfk} \kk_{\gi}^{\sfl} \, \bigl[(\tilde{K}^{\sfk}{}_{\sfl})^{\rmT}\,\hat{G} + \hat{G}\,\tilde{K}^{\sfk}{}_{\sfl}\bigr]_{IJ}\,.
\end{align}
Combining this with Eq.~\eqref{eq:delta-bP}, we can easily see that the combination
\begin{align}
 \hat{G}_{IJ}\,\hat{\bm{\cP}}^I\wedge *_\gamma\hat{\bm{\cP}}^J = \cM_{IJ}\,\bm{\cP}^I\wedge *_\gamma\bm{\cP}^J\qquad \bigl(\bm{\cP}^I\equiv (L^{-1})^I{}_J\,\hat{\bm{\cP}}^J\equiv P^I + \KK_{\gi}{}^I \cA^{\gi}\bigr)\,,
\end{align}
is gauge invariant. 
We thus use $\cM_{IJ}(x)\,\bm{\cP}^I\wedge *_\gamma\bm{\cP}^J$ as the kinetic term of our gauged action. 

In addition, Eq.~\eqref{eq:delta-bP} indicates that, if we contract the curved indices of an arbitrary component of $\hat{\bm{\cP}}^I$ with $Dx^{\sfi}$\,, we can construct a gauge-invariant combination. 
For example, in M-theory, 
\begin{align}
 \hat{\bm{\cP}}_{1+2}\equiv \hat{\bm{\cP}}_{i_2}\wedge Dx^{i_2} \,,\qquad \hat{\bm{\cP}}_{1+5}\equiv\hat{\bm{\cP}}_{i_5}\wedge Dx^{i_5} \,,\quad\cdots\,,
\end{align}
are invariant.
In type IIB theory,
\begin{align}
 \hat{\bm{\cP}}^{\Ba}_{1+1}\equiv\hat{\bm{\cP}}^{\Ba}_{m}\wedge Dx^{m} \,,\qquad 
 \hat{\bm{\cP}}_{1+3}\equiv\hat{\bm{\cP}}_{m_3}\wedge Dx^{m_3}\,,\qquad
 \hat{\bm{\cP}}^{\Ba}_{1+5}\equiv\hat{\bm{\cP}}^{\Ba}_{m_5}\wedge Dx^{m_5} \,,\quad\cdots\,,
\end{align}
are invariant. 
In $\Omega_{\text{brane}}$\,, the auxiliary fields $\hat{P}^{\cdots}_{\cdots}$ always appear with the combination, such as $\hat{P}_{i_2}\wedge \rmd x^{i_2}$, and thus, when we construct the gauged action, we replace these combinations with the above gauge-invariant combinations. 

To construct the gauge-invariant actions, we further need to find some gauge-invariant combinations that only involve the supergravity potentials, the worldvolume gauge fields, the generalized Killing vector fields $\KK_{\gi}{}^I$, and the gauge fields $\cA^{\gi}$\,. 
Unfortunately, we do not find a systematic way of finding such combinations, and we find these on a case-by-case basis. 
It turns out that the gauge transformation rules for the worldvolume gauge fields can be determined successively from lower-order ones.

\subsection{Gauged membrane action}

Let us begin by considering the membrane action. 
Before the gauging, the Lagrangian for the membrane theory is given by
\begin{align}
 \cL_{(\text{M2})} = -\tfrac{1}{6}\,\cM_{IJ}\, P^I\wedge *_\gamma P^J -\tfrac{\mu_2}{3}\,\hat{P}_{i_2}\wedge \rmd x^{i_2} + \mu_2\,\bigl(C_3 -F_2\bigr) \,,
\end{align}
where we have absorbed the scalar field $\lambda$ into the intrinsic metric $\gamma$. 
The first two terms can be replaced by the gauge-invariant combinations
\begin{align}
 \hat{\cL}_0 \equiv - \tfrac{1}{6}\,\cM_{IJ}\, \bm{\cP}^I\wedge *_\gamma \bm{\cP}^J - \tfrac{\mu_2}{3}\, \hat{\bm{\cP}}_{1+2} \,.
\end{align}
The non-trivial part is the WZ term. 
Since $C_{i_3}$ transforms as a function,
\begin{align}
 \delta C_{i_3} = \epsilon^{\gi}\, \kk_{\gi}^j\,\partial_j C_{i_3}\,,
\end{align}
by using the generalized Killing equation \eqref{eq:M-gKE} and $\delta \rmd x^i=\epsilon^{\gi}\,\partial_j \kk_{\gi}^i\,\rmd x^j + \rmd\epsilon^{\gi}\,\kk_{\gi}^i$\,, we find
\begin{align}
 \delta C_3 = -\epsilon^{\gi}\,\rmd \tilde{\kk}_{\gi}^{(2)}
 + \rmd \epsilon^{\gi} \wedge \iota_{\gi}C_3\,.
\label{eq:delta-C3}
\end{align}
Then, by requiring
\begin{align}
 \delta A_2 = -\epsilon^{\gi}\, \tilde{\kk}_{\gi}^{(2)}\,,
\end{align}
the variation of the WZ term is simplified as
\begin{align}
 \delta (C_3-F_3) = \rmd\epsilon^{\gi}\wedge \hat{\kk}_{\gi}^{(2)}\,.
\end{align}
To compensate this variation, we take an approach similar to the one by Hull and Spence \cite{Hull:1989jk,Hull:1990ms}. 
We can show the identities
\begin{align}
\begin{split}
 \delta \bigl(\cA^{\gi}\wedge \hat{\kk}_{\gi}^{(2)}\bigr) &= -\rmd\epsilon^{\gi}\wedge\hat{\kk}_{\gi}^{(2)} + \cA^{\gi_1}\wedge \rmd \epsilon^{\gi_2}\wedge\iota_{\gi_2}\hat{\kk}_{\gi_1}^{(2)}\,,
\\
 \delta \bigl(\tfrac{1}{2}\,\cA^{\gi_1\gi_2}\wedge \iota_{\gi_2}\hat{\kk}_{\gi_1}^{(2)}\bigr) 
 &= - \cA^{\gi_1}\wedge \rmd \epsilon^{\gi_2}\wedge\iota_{[\gi_2}\hat{\kk}_{\gi_1]}^{(2)}
 + \tfrac{1}{2}\,\cA^{\gi_1\gi_2}\wedge\rmd \epsilon^{\gi_3}\wedge\iota_{\gi_3\gi_2}\hat{\kk}_{\gi_1}^{(2)}
\,,
\\
 \delta \bigl(\tfrac{1}{3!}\,\cA^{\gi_1\gi_2\gi_3}\wedge \iota_{\gi_3\gi_2}\hat{\kk}_{\gi_1}^{(2)}\bigr) &= 
 -\tfrac{1}{2}\,\cA^{\gi_1\gi_2}\wedge \rmd \epsilon^{\gi_3}\wedge \iota_{[\gi_3\gi_2}\hat{\kk}_{\gi_1]}^{(2)} \,,
\end{split}
\end{align}
and by using Eq.~\eqref{eq:M2-SC}, we find
\begin{align}
 \delta \bigl(\cA^{\gi}\wedge \hat{\kk}_{\gi}^{(2)}+\tfrac{1}{2}\,\cA^{\gi_1\gi_2}\wedge \iota_{\gi_2}\hat{\kk}_{\gi_1}^{(2)}+\tfrac{1}{3!}\,\cA^{\gi_1\gi_2\gi_3}\wedge \iota_{\gi_3\gi_2}\hat{\kk}_{\gi_1}^{(2)}\bigr) &= -\rmd\epsilon^{\gi}\wedge\hat{\kk}_{\gi}^{(2)}\,.
\end{align}
Then the gauge-invariant extension of the WZ term is
\begin{align}
 \hat{\cL}^{\text{WZ}}_{(\text{M2})} = \mu_2\,\bigl(C_3-F_3 + \cA^{\gi}\wedge \hat{\kk}_{\gi}^{(2)}+\tfrac{1}{2}\,\cA^{\gi_1\gi_2}\wedge \iota_{\gi_2}\hat{\kk}_{\gi_1}^{(2)}+\tfrac{1}{3!}\,\cA^{\gi_1\gi_2\gi_3}\wedge \iota_{\gi_3\gi_2}\hat{\kk}_{\gi_1}^{(2)}\bigr) \,.
\end{align}
Considering the sum $\hat{\cL}_{(\text{M2})}\equiv \hat{\cL}_0+\hat{\cL}^{\text{WZ}}_{(\text{M2})}$\,, we obtain the gauged membrane action
\begin{align}
 \hat{S}_{(\text{M2})}&= -\int_{\Sigma_3} \bigl(\tfrac{1}{6}\,\cM_{IJ}\,\bm{\cP}^I\wedge *_\gamma\bm{\cP}^J + \tfrac{\mu_2}{3}\,\hat{\bm{\cP}}_{1+2} \bigr) + \int_{\Sigma_3} \hat{\cL}^{\text{WZ}}_{(\text{M2})}
\\
 &= - \int_{\Sigma_3} \bigl(\tfrac{1}{6}\,\cM_{IJ}\,\bm{\cP}^I\wedge *_\gamma\bm{\cP}^J + \tfrac{\mu_2}{3}\,\bm{\cP}_{i_2}\wedge D x^{i_2} \bigr) - \mu_2\int_{\Sigma_3}F_3
\nn\\
 &\quad + \mu_2\int_{\Sigma_3} \bigl(\cA^{\gi}\wedge \tilde{\kk}^{(2)}_{\gi} + \tfrac{1}{2!}\,\cA^{\gi_1\gi_2}\wedge \iota_{\gi_2}\tilde{\kk}^{(2)}_{\gi_1} + \tfrac{1}{3!}\,\cA^{\gi_1\gi_2\gi_3}\,\iota_{\gi_3}\iota_{\gi_2}\tilde{\kk}^{(2)}_{\gi_1} \bigr)\,.
\end{align}

\subsection{Gauged M5-brane action}

Before the gauging, the Lagrangian for an M5-brane is
\begin{align}
 \cL_{(\text{M5})} = \tfrac{1}{12}\,\cM_{IJ} \,P^I\wedge *_{\gamma} P^J - \tfrac{\mu_5}{6}\,\bigl[\hat{P}_{i_5}\wedge \rmd x^{i_5} + \hat{P}_{i_2}\wedge D x^{i_2}\wedge \bigl(C_3-F_3\bigr)\bigr] + \cL^{\text{WZ}}_{(\text{M5})} \,,
\end{align}
where again we have absorbed $\lambda$ into $\gamma$ and
\begin{align}
 \cL^{\text{WZ}}_{(\text{M5})}
 = \mu_5\,\bigl(C_6-\tfrac{1}{2}\,C_3\wedge F_3 - F_6 \bigr)\,.
\end{align}
Except the WZ terms, we can find the gauge-invariant extension as
\begin{align}
 \hat{\cL}_0 &= \tfrac{1}{12} \,\cM_{IJ} \,\bm{\cP}^I\wedge *_{\gamma} \bm{\cP}^J 
\\
 &\quad - \tfrac{\mu_5}{6}\,\bigl[\hat{\bm{\cP}}_{1+5} + \hat{\bm{\cP}}_{1+2}\wedge \bigl(C_3-F_3 + \cA^{\gi}\wedge \hat{\kk}^{(2)}_{\gi} + \tfrac{1}{2!}\,\cA^{\gi_1\gi_2}\wedge \iota_{\gi_2}\hat{\kk}^{(2)}_{\gi_1} + \tfrac{1}{3!}\,\cA^{\gi_1\gi_2\gi_3}\wedge \iota_{\gi_3\gi_2}\hat{\kk}^{(2)}_{\gi_1} \bigr)\bigr] \,.
\nn
\end{align}
The variation of the WZ term becomes
\begin{align}
 \mu_5^{-1}\,\delta \cL^{\text{WZ}}_{(\text{M5})}
 =\rmd \epsilon^{\gi}\wedge\bigl[\hat{\kk}_{\gi}^{(5)}
 + \tfrac{1}{2}\,(C_3-F_3)\wedge \hat{\kk}_{\gi}^{(2)} \bigr]
 +\rmd \bigl(-\epsilon^{\gi}\,\tilde{\kk}_{\gi}^{(5)} 
 + \tfrac{1}{2}\, \epsilon^{\gi}\,\tilde{\kk}_{\gi}^{(2)} \wedge F_3\bigr)- \delta F_6\,,
\label{eq:delta-WZ-M5}
\end{align}
where we have used
\begin{align}
 \delta C_6 = -\epsilon^{\gi}\,\rmd \tilde{\kk}_{\gi}^{(5)} - \tfrac{1}{2}\,\epsilon^{\gi}\,C_3\wedge \rmd \tilde{\kk}_{\gi}^{(2)} 
 + \rmd \epsilon^{\gi}\wedge \iota_{\gi}C_6\,.
\end{align}
Then, by requiring
\begin{align}
 \delta A_5= -\epsilon^{\gi}\,\tilde{\kk}_{\gi}^{(5)} 
 + \tfrac{1}{2}\, \epsilon^{\gi}\,\tilde{\kk}_{\gi}^{(2)} \wedge F_3\,,
\end{align}
the variation \eqref{eq:delta-WZ-M5} is simplified as
\begin{align}
 \mu_5^{-1}\,\delta \cL^{\text{WZ}}_{(\text{M5})}
 &=\rmd \epsilon^{\gi}\wedge\bigl[\hat{\kk}_{\gi}^{(5)}
 + \tfrac{1}{2}\,(C_3-F_3)\wedge \hat{\kk}_{\gi}^{(2)} \bigr]\,.
\end{align}

Again we find a non-trivial combination which compensate this variation. 
To compensate the first term (i.e., $\rmd \epsilon^{\gi}\wedge \hat{\kk}_{\gi}^{(5)}$), it is useful to introduce the Lagrangian
\begin{align}
 \cL_1 \equiv \mu_5\,\bigl(\cA^{\gi}\wedge \hat{\kk}_{\gi}^{(5)}+\tfrac{1}{2!}\,\cA^{\gi_1\gi_2}\wedge \iota_{\gi_2}\hat{\kk}_{\gi_1}^{(5)}+\cdots +\tfrac{1}{6!}\,\cA^{\gi_1\cdots \gi_6}\wedge \iota_{\gi_6\cdots \gi_2} \hat{\kk}_{\gi_1}^{(5)}\bigr)\,,
\end{align}
and use the identities:
\begin{align}
 \delta \bigl(\cA^{\gi}\wedge \hat{\kk}_{\gi}^{(5)}\bigr) &= -\rmd\epsilon^{\gi}\wedge\hat{\kk}_{\gi}^{(5)} + \cA^{\gi_1}\wedge \rmd \epsilon^{\gi_2}\wedge\iota_{\gi_2}\hat{\kk}_{\gi_1}^{(5)}\,,
\nn\\
 \delta \bigl(\tfrac{1}{2!}\,\cA^{\gi_1\gi_2}\wedge \iota_{\gi_2}\hat{\kk}_{\gi_1}^{(5)}\bigr) 
 &= - \cA^{\gi_1}\wedge \rmd \epsilon^{\gi_2}\wedge\iota_{[\gi_2}\hat{\kk}_{\gi_1]}^{(5)}
 + \tfrac{1}{2!}\,\cA^{\gi_1\gi_2}\wedge\rmd \epsilon^{\gi_3}\wedge\iota_{\gi_3\gi_2}\hat{\kk}_{\gi_1}^{(5)}
\,,
\nn\\
 \delta \bigl(\tfrac{1}{3!}\,\cA^{\gi_1\gi_2\gi_3}\wedge \iota_{\gi_3\gi_2}\hat{\kk}_{\gi_1}^{(5)}\bigr) &= 
 -\tfrac{1}{2}\,\cA^{\gi_1\gi_2}\wedge \rmd \epsilon^{\gi_3}\wedge \iota_{[\gi_3\gi_2}\hat{\kk}_{\gi_1]}^{(5)}
 + \tfrac{1}{3!}\,\cA^{\gi_1\gi_2\gi_3}\wedge \rmd \epsilon^{\gi_4}\wedge\iota_{\gi_4\gi_3\gi_2}\hat{\kk}_{\gi_1}^{(5)}\,,
\nn\\
 \delta \bigl(\tfrac{1}{4!}\,\cA^{\gi_1\cdots \gi_4}\wedge \iota_{\gi_4\cdots \gi_2} \hat{\kk}_{\gi_1}^{(5)}\bigr) &= 
 -\tfrac{1}{3!}\,\cA^{\gi_1\gi_2\gi_3}\wedge \rmd \epsilon^{\gi_4}\wedge \iota_{[\gi_4\gi_3\gi_2}\hat{\kk}_{\gi_1]}^{(5)}
 + \tfrac{1}{4!}\,\cA^{\gi_1\cdots\gi_4}\wedge \rmd \epsilon^{\gi_5}\wedge\iota_{\gi_5\cdots \gi_2}\hat{\kk}_{\gi_1}^{(5)}\,,
\nn\\
 \delta \bigl(\tfrac{1}{5!}\,\cA^{\gi_1\cdots \gi_5}\wedge \iota_{\gi_5\cdots \gi_2} \hat{\kk}_{\gi_1}^{(5)}\bigr) &= 
 -\tfrac{1}{4!}\,\cA^{\gi_1\cdots \gi_4}\wedge \rmd \epsilon^{\gi_5}\wedge \iota_{[\gi_5\cdots\gi_2}\hat{\kk}_{\gi_1]}^{(5)}
 + \tfrac{1}{5!}\,\cA^{\gi_1\cdots\gi_5}\wedge \rmd \epsilon^{\gi_6}\wedge\iota_{\gi_6\cdots \gi_2}\hat{\kk}_{\gi_1}^{(5)}\,,
\nn\\
 \delta \bigl(\tfrac{1}{6!}\,\cA^{\gi_1\cdots \gi_6}\wedge \iota_{\gi_6\cdots \gi_2} \hat{\kk}_{\gi_1}^{(5)}\bigr) &= 
 -\tfrac{1}{5!}\,\cA^{\gi_1\cdots \gi_5}\wedge \rmd \epsilon^{\gi_6}\wedge \iota_{[\gi_6\cdots\gi_2}\hat{\kk}_{\gi_1]}^{(5)}\,.
\end{align}
To compensate the second term, it is useful to consider
\begin{align}
 \cL_2= -\tfrac{\mu_5}{2}\,(C_3-F_3)\wedge \bigl(\cA^{\gi}\wedge \hat{\kk}^{(2)}_{\gi} + \tfrac{1}{2!}\,\cA^{\gi_1\gi_2}\wedge \iota_{\gi_2}\hat{\kk}^{(2)}_{\gi_1} +\tfrac{1}{3!}\,\cA^{\gi_1\gi_2\gi_3}\,\iota_{\gi_3\gi_2}\hat{\kk}^{(2)}_{\gi_1}\bigr)\,,
\end{align}
whose variation is
\begin{align}
\begin{split}
 \mu_5^{-1}\,\delta \cL_2&=-\tfrac{1}{2}\,\rmd\epsilon^{\gi}\wedge (C_3-F_3)\wedge \hat{\kk}_{\gi}^{(2)}
 +\tfrac{1}{2}\,\cA^{\gi_1}\wedge\rmd\epsilon^{\gi_2}\wedge \hat{\kk}_{\gi_2}^{(2)}\wedge \hat{\kk}^{(2)}_{\gi_1} 
\\
 &\quad -\tfrac{1}{4}\,\cA^{\gi_1\gi_2}\wedge \rmd\epsilon^{\gi_3}\wedge \hat{\kk}_{\gi_3}^{(2)}\wedge \iota_{\gi_2}\hat{\kk}^{(2)}_{\gi_1}
 + \tfrac{1}{2\cdot 3!}\,\cA^{\gi_1\gi_2\gi_3}\wedge \rmd\epsilon^{\gi_4} \, \iota_{\gi_3\gi_2}\hat{\kk}^{(2)}_{\gi_1} \, \hat{\kk}_{\gi_4}^{(2)}\,.
\end{split}
\end{align}

We compute $\delta (\cL_1+\cL_2)$ and find that the terms which are linear in $\cA^{\gi}$ are canceled out by using
\begin{align}
 \cA^{\gi_1}\wedge \rmd \epsilon^{\gi_2}\wedge\bigl(\iota_{(\gi_1}\hat{\kk}_{\gi_2)}^{(5)}+ \tfrac{1}{2}\,\hat{\kk}^{(2)}_{\gi_1}\wedge\hat{\kk}^{(2)}_{\gi_2}\bigr)=0\,,
\end{align}
which follows from the isotropy condition \eqref{eq:M5-SC}. 
However, we find that terms that are quadratic in $\cA^{\gi}$ are not canceled out. 
To compensate the variation we add the term
\begin{align}
 \cL_3 \equiv -\tfrac{\mu_5}{12}\,\cA^{\gi_1\gi_2\gi_3}\wedge \iota_{\gi_3}\hat{\kk}^{(2)}_{\gi_2}\wedge\hat{\kk}^{(2)}_{\gi_1}\,,
\end{align}
whose variation is
\begin{align}
 \mu_5^{-1}\,\delta \cL_3 
 &= \tfrac{1}{4}\,\cA^{\gi_1\gi_2}\wedge \rmd\epsilon^{\gi_3}\wedge \iota_{[\gi_3}\hat{\kk}^{(2)}_{\gi_2}\wedge\hat{\kk}^{(2)}_{\gi_1]}
\nn\\
 &\quad + \tfrac{1}{4!}\,\cA^{\gi_1\gi_2\gi_3}\wedge\rmd \epsilon^{\gi_4}\wedge \bigl(-2\,\iota_{\gi_4\gi_3}\hat{\kk}^{(2)}_{\gi_2}\wedge\hat{\kk}^{(2)}_{\gi_1}+2\,\iota_{\gi_3}\hat{\kk}^{(2)}_{\gi_2}\wedge\iota_{\gi_4}\hat{\kk}^{(2)}_{\gi_1}\bigr)\,.
\end{align}
Then, using the identity
\begin{align}
 \iota_{\gi_3\gi_2}\hat{\kk}^{(5)}_{\gi_1}-\iota_{\gi_2\gi_1}\hat{\kk}^{(5)}_{\gi_3}
 - \iota_{\gi_2}\hat{\kk}^{(2)}_{\gi_3}\wedge\hat{\kk}^{(2)}_{\gi_1} 
 - \iota_{\gi_2}\hat{\kk}^{(2)}_{\gi_1}\wedge\hat{\kk}^{(2)}_{\gi_3} = 0 \,,
\end{align}
which follows from \eqref{eq:M5-SC} and gives
\begin{align}
 0&=\tfrac{1}{3!}\,\cA^{\gi_1\gi_2}\wedge\rmd\epsilon^{\gi_3}\wedge\bigl(\iota_{\gi_3\gi_2}\hat{\kk}^{(5)}_{\gi_1}-\iota_{\gi_2\gi_1}\hat{\kk}^{(5)}_{\gi_3}
 - \iota_{\gi_2}\hat{\kk}^{(2)}_{\gi_3}\wedge\hat{\kk}^{(2)}_{\gi_1} 
 - \iota_{\gi_2}\hat{\kk}^{(2)}_{\gi_1}\wedge\hat{\kk}^{(2)}_{\gi_3}\bigr)
\nn\\
 &=\tfrac{1}{3!}\,\cA^{\gi_1\gi_2}\wedge\rmd\epsilon^{\gi_3}\wedge\bigl(\iota_{\gi_3\gi_2}\hat{\kk}^{(5)}_{\gi_1}-\iota_{\gi_2\gi_1}\hat{\kk}^{(5)}_{\gi_3}\bigr)
 -\tfrac{1}{4}\,\cA^{\gi_1\gi_2}\wedge\rmd\epsilon^{\gi_3}\wedge\hat{\kk}^{(2)}_{\gi_3} \wedge \iota_{\gi_2}\hat{\kk}^{(2)}_{\gi_1}
\nn\\
 &\quad + \tfrac{1}{4}\,\cA^{\gi_1\gi_2}\wedge\rmd\epsilon^{\gi_3}\wedge \iota_{[\gi_3}\hat{\kk}^{(2)}_{\gi_2}\wedge\hat{\kk}^{(2)}_{\gi_1]} \,,
\end{align}
we find that the quadratic terms are canceled out. 

We repeat this procedure. 
We introduce
\begin{align}
 \cL_4 \equiv -\tfrac{\mu_5}{4!}\,\cA^{\gi_1\cdots \gi_4}\wedge \iota_{\gi_4\gi_3}\hat{\kk}^{(2)}_{\gi_2}\wedge\hat{\kk}^{(2)}_{\gi_1}\,,
\end{align}
whose variation is
\begin{align}
 \mu_5^{-1}\,\delta \cL_4
 = \tfrac{1}{3!}\,\cA^{\gi_1\gi_2\gi_3}\wedge\rmd \epsilon^{\gi_4}\wedge \iota_{[\gi_4\gi_3}\hat{\kk}^{(2)}_{\gi_2}\,\hat{\kk}^{(2)}_{\gi_1]}
 + \tfrac{1}{4!}\,\cA^{\gi_1\cdots\gi_4}\wedge\rmd\epsilon^{\gi_5}\wedge\iota_{\gi_4\gi_3}\hat{\kk}^{(2)}_{\gi_2}\,\iota_{\gi_1}\hat{\kk}^{(2)}_{\gi_5}\,.
\end{align}
The identity
\begin{align}
\begin{split}
 &\iota_{\gi_4\gi_3\gi_2}\hat{\kk}^{(5)}_{\gi_1}
 + \iota_{\gi_3\gi_2\gi_1}\hat{\kk}^{(5)}_{\gi_4}
\\
 &- \iota_{\gi_2\gi_3}\hat{\kk}^{(2)}_{\gi_4}\,\hat{\kk}^{(2)}_{\gi_1} 
 - \iota_{\gi_2\gi_3}\hat{\kk}^{(2)}_{\gi_1}\,\hat{\kk}^{(2)}_{\gi_4} 
 + \iota_{\gi_3}\hat{\kk}^{(2)}_{\gi_1}\wedge\iota_{\gi_2}\hat{\kk}^{(2)}_{\gi_4}
 + \iota_{\gi_3}\hat{\kk}^{(2)}_{\gi_4}\wedge\iota_{\gi_2}\hat{\kk}^{(2)}_{\gi_1} = 0\,,
\end{split}
\end{align}
gives
\begin{align}
 0&=\tfrac{1}{4!}\,\cA^{\gi_1\gi_2\gi_3}\wedge\rmd\epsilon^{\gi_4}\wedge\bigl(\iota_{\gi_4\gi_3\gi_2}\hat{\kk}^{(5)}_{\gi_1}
 + \iota_{\gi_3\gi_2\gi_1}\hat{\kk}^{(5)}_{\gi_4}
 - \iota_{\gi_2\gi_3}\hat{\kk}^{(2)}_{\gi_4}\,\hat{\kk}^{(2)}_{\gi_1} 
 - \iota_{\gi_2\gi_3}\hat{\kk}^{(2)}_{\gi_1}\,\hat{\kk}^{(2)}_{\gi_4} 
 + 2\,\iota_{\gi_3}\hat{\kk}^{(2)}_{\gi_1}\wedge\iota_{\gi_2}\hat{\kk}^{(2)}_{\gi_4} \bigr)
\nn\\
 &=\tfrac{1}{4!}\,\cA^{\gi_1\gi_2\gi_3}\wedge\rmd\epsilon^{\gi_4}\wedge\bigl(\iota_{\gi_4\gi_3\gi_2}\hat{\kk}^{(5)}_{\gi_1}
 + \iota_{\gi_3\gi_2\gi_1}\hat{\kk}^{(5)}_{\gi_4} \bigr)
 +\tfrac{1}{2\cdot 3!}\,\cA^{\gi_1\gi_2\gi_3}\wedge\rmd\epsilon^{\gi_4}\wedge
 \iota_{\gi_3\gi_2}\hat{\kk}^{(2)}_{\gi_1}\,\hat{\kk}^{(2)}_{\gi_4} 
\nn\\
 &+\tfrac{1}{4!}\,\cA^{\gi_1\gi_2\gi_3}\wedge\rmd\epsilon^{\gi_4}\wedge\bigl(
 -2\,\iota_{\gi_4\gi_3}\hat{\kk}^{(2)}_{\gi_2}\,\hat{\kk}^{(2)}_{\gi_1}
 + 2\,\iota_{\gi_3}\hat{\kk}^{(2)}_{\gi_2}\wedge\iota_{\gi_4}\hat{\kk}^{(2)}_{\gi_1} \bigr)
\nn\\
 &+\tfrac{1}{3!}\,\cA^{\gi_2\gi_3\gi_1}\wedge\rmd\epsilon^{\gi_4}\,\iota_{[\gi_4\gi_1}\hat{\kk}^{(2)}_{\gi_3}\wedge\hat{\kk}^{(2)}_{\gi_2]} \,,
\end{align}
and then we find that the cubic terms are also canceled out. 
We further consider
\begin{align}
 \cL_5 \equiv \tfrac{\mu_5}{5!}\,\cA^{\gi_1\cdots\gi_5} \wedge \iota_{\gi_5\gi_4}\hat{\kk}^{(2)}_{\gi_3}\wedge\iota_{\gi_2}\hat{\kk}^{(2)}_{\gi_1}\,,
\end{align}
which satisfy
\begin{align}
 \mu_5^{-1}\,\delta \cL_5
 =- \tfrac{1}{4!}\,\cA^{\gi_1\cdots\gi_4}\wedge\rmd\epsilon^{\gi_5}\wedge \iota_{[\gi_5\gi_4}\hat{\kk}^{(2)}_{\gi_3}\,\iota_{\gi_2}\hat{\kk}^{(2)}_{\gi_1]}
 +\tfrac{1}{5!}\,\cA^{\gi_1\cdots\gi_5} \wedge\rmd\epsilon^{\gi_6}\, \iota_{\gi_5\gi_4}\hat{\kk}^{(2)}_{\gi_3}\,\iota_{\gi_2\gi_1}\hat{\kk}^{(2)}_{\gi_6} \,.
\end{align}
Using the identities
\begin{align}
 0&=\tfrac{1}{5!}\,\cA^{\gi_1\cdots \gi_4}\wedge\rmd\epsilon^{\gi_5}\wedge\bigl(
 \iota_{\gi_5\cdots \gi_2}\hat{\kk}^{(5)}_{\gi_1}
 -\iota_{\gi_4\cdots\gi_1}\hat{\kk}^{(5)}_{\gi_5}\bigr)
\nn\\
 &\quad + \tfrac{1}{4!}\,\cA^{\gi_1\cdots \gi_4}\wedge\rmd\epsilon^{\gi_5}\wedge \iota_{\gi_4\gi_3}\hat{\kk}^{(2)}_{\gi_2}\wedge\iota_{\gi_1}\hat{\kk}^{(2)}_{\gi_5} 
 - \tfrac{1}{4!}\,\cA^{\gi_1\cdots \gi_4}\wedge\rmd\epsilon^{\gi_5}\wedge \iota_{[\gi_5\gi_4}\hat{\kk}^{(2)}_{\gi_3}\wedge\iota_{\gi_2}\hat{\kk}^{(2)}_{\gi_1]} \,,
\\
 0&=\tfrac{1}{6!}\,\cA^{\gi_1\cdots \gi_5}\wedge\rmd\epsilon^{\gi_6}\wedge\bigl(
 \iota_{\gi_6\cdots\gi_2}\hat{\kk}^{(5)}_{\gi_1}
 +\iota_{\gi_5\cdots \gi_1}\hat{\kk}^{(5)}_{\gi_6}\bigr)
\nn\\
 &\quad + \tfrac{1}{5!}\,\cA^{\gi_1\cdots \gi_5}\wedge\rmd\epsilon^{\gi_6}\wedge \iota_{\gi_5\gi_4}\hat{\kk}^{(2)}_{\gi_3}\,\iota_{\gi_2\gi_1}\hat{\kk}^{(2)}_{\gi_6} \,,
\end{align}
we find that
\begin{align}
 \hat{\cL}^{\text{WZ}}_{(\text{M5})} 
 \equiv \cL^{\text{WZ}}_{(\text{M5})} + \cL_1 + \cdots + \cL_5\,,
\end{align}
is gauge invariant.
Then the Lagrangian for the gauged M5-brane theory is $\hat{\cL}_{(\text{M5})}\equiv \hat{\cL}_0+\hat{\cL}^{\text{WZ}}_{(\text{M5})}$\,. 

In summary, the gauged M5-brane action is given by
\begin{align}
 \hat{S}_{(\text{M5})}&=\tfrac{1}{12} \int_{\Sigma_6}\,\cM_{IJ} \,\bm{\cP}^I\wedge *_{\gamma} \bm{\cP}^J 
 +\mu_5\int_{\Sigma_6}\bigl(C_6-\tfrac{1}{2}\,C_3\wedge F_3 - F_6 \bigr)
\\
 &\quad +\mu_5\int_{\Sigma_6}\bigl(\cA^{\gi}\wedge \hat{\kk}_{\gi}^{(5)}+\tfrac{1}{2!}\,\cA^{\gi_1\gi_2}\wedge \iota_{\gi_2}\hat{\kk}_{\gi_1}^{(5)}+\cdots +\tfrac{1}{6!}\,\cA^{\gi_1\cdots \gi_6}\wedge \iota_{\gi_6\cdots \gi_2} \hat{\kk}_{\gi_1}^{(5)}- \tfrac{1}{6}\,\hat{\bm{\cP}}_{1+5}\bigr)
\nn\\
 &\quad -\tfrac{\mu_5}{2}\int_{\Sigma_6}(C_3-F_3+\tfrac{1}{3}\,\hat{\bm{\cP}}_{1+2})
\nn\\
 &\quad\qquad\qquad \wedge \bigl(\cA^{\gi}\wedge \hat{\kk}^{(2)}_{\gi} + \tfrac{1}{2!}\,\cA^{\gi_1\gi_2}\wedge \iota_{\gi_2}\hat{\kk}^{(2)}_{\gi_1} + \tfrac{1}{3!}\,\cA^{\gi_1\gi_2\gi_3}\,\iota_{\gi_3}\iota_{\gi_2}\hat{\kk}^{(2)}_{\gi_1}- \tfrac{1}{3}\,\hat{\bm{\cP}}_{1+2}\bigr)
\nn\\
 &\quad -\tfrac{\mu_5}{2}\int_{\Sigma_6}\bigl(\tfrac{1}{3!}\,\cA^{\gi_1\gi_2\gi_3}\wedge \iota_{\gi_3}\hat{\kk}^{(2)}_{\gi_2}\wedge\hat{\kk}^{(2)}_{\gi_1} 
 +\tfrac{2}{4!}\,\cA^{\gi_1\cdots\gi_4}\wedge \iota_{\gi_4\gi_3}\hat{\kk}^{(2)}_{\gi_2}\,\hat{\kk}^{(2)}_{\gi_1}
\nn\\
 &\quad\qquad\qquad -\tfrac{2}{5!}\,\cA^{\gi_1\cdots\gi_5} \, \iota_{\gi_5\gi_4}\hat{\kk}^{(2)}_{\gi_3}\,\iota_{\gi_2}\hat{\kk}^{(2)}_{\gi_1}\bigr)\,.\nn
\end{align}
We note that we have repeatedly used the identities that follow from the isotropy condition of the gauge algebra $\mathfrak{f}$. 
As long as the isotropy, which is a $U$-duality-invariant condition, is assumed, we can construct the gauged action without requiring any further conditions on the generalized Killing vector fields. 

In M-theory, the next brane we can consider is the Kaluza--Klein monopole (KKM). 
The ungauged action of a KKM has been studied partially in \cite{1712.10316}. 
We can also consider various exotic branes in M-theory, such as the $5^3$-brane or the $2^6$-brane. 
The worldvolume theories of these exotic branes are very involved and here we do not consider these. 
Here we only point out an interesting open question associated with these exotic branes. 

Exotic branes always require the existence of several isometry directions.
For KKM, a single direction is required, and the $5^3$-brane and the $2^6$-brane require three and six isometry directions, respectively. 
Usually, these isometries are assumed to be Abelian, but here we have non-Abelian isometries generated by $\KK_{\gi}{}^I$. 
If we could find the gauged action for an exotic brane by gauging the non-Abelian isometries, we will find novel objects, which we call non-Abelian exotic branes. 
We believe that this would be an interesting direction to explore in the future. 

\subsection{Gauged $(p,q)$-string action}

Now, let us turn to the type IIB branes. 
The ungauged Lagrangian for a $(p,q)$-string is
\begin{align}
 \cL_{(p,q)\text{-1}} = \tfrac{1}{4} \Exp{\lambda}\cM_{IJ} \,P^I\wedge *_{\gamma} P^J 
 -\tfrac{\mu_1}{2}\,q_{\Ba}\,\hat{P}^{\Ba}_{m} \wedge \rmd x^m 
 + \mu_1\,q_{\Ba}\,(B^{\Ba}_2-F^{\Ba}_2) \,.
\end{align}
The first two terms can be extended to the gauge-invariant ones
\begin{align}
 \hat{\cL}_0 = \tfrac{1}{4} \Exp{\lambda}\cM_{IJ} \,\bm{\cP}^I\wedge *_{\gamma} \bm{\cP}^J 
 -\tfrac{\mu_1}{2}\,q_{\Ba}\,\hat{\bm{\cP}}^{\Ba}_{1+1} \,. 
\end{align}
Using
\begin{align}
 \delta B^{\Ba}_2 = -\epsilon^{\gi}\,\rmd \tilde{\kk}_{\gi}^{\Ba(1)}
 + \rmd \epsilon^{\gi} \wedge \iota_{\gi}B^{\Ba}_2\,,
\end{align}
and assuming
\begin{align}
 \delta A_1^{\Ba} = -\epsilon^{\gi}\, \tilde{\kk}_{\gi}^{\Ba(1)}\,,
\end{align}
we find
\begin{align}
 \delta \bigl(B^{\Ba}_2-F^{\Ba}_2\bigr) = \rmd \epsilon^{\gi}\wedge \tilde{\kk}_{\gi}^{\Ba(1)}\,.
\end{align}
Similar to the membrane case, by using the isotropy condition \eqref{eq:B1-SC}, we can show
\begin{align}
 \delta \bigl( \cA^{\gi}\wedge \hat{\kk}_{\gi}^{\Ba(1)}+\tfrac{1}{2!}\,\cA^{\gi_1\gi_2}\,\iota_{\gi_2}\hat{\kk}_{\gi_1}^{\Ba(1)} \bigr)
 = -\rmd \epsilon^{\gi}\wedge \tilde{\kk}_{\gi}^{\Ba(1)}\,.
\end{align}
We then obtain the gauge-invariant extension of the WZ term as
\begin{align}
 \hat{\cL}^{\text{WZ}}_{(p,q)\text{-1}} \equiv \mu_1\,q_{\Ba}\,\bigl(B^{\Ba}_2-F^{\Ba}_2+\cA^{\gi}\wedge \hat{\kk}_{\gi}^{\Ba(1)}+\tfrac{1}{2!}\,\cA^{\gi_1\gi_2}\,\iota_{\gi_2}\hat{\kk}_{\gi_1}^{\Ba(1)}\bigr)\,.
\end{align}
Combining these two terms as $\hat{\cL}_{(p,q)\text{-1}}\equiv \hat{\cL}_0 + \hat{\cL}^{\text{WZ}}_{(p,q)\text{-1}}$\,, we obtain the gauged action for a $(p,q)$-string
\begin{align}
 S&= \frac{1}{4}\int_{\Sigma_2} \Exp{\lambda}\cM_{IJ}\,\bm{\cP}^I\wedge *_\gamma\bm{\cP}^J 
 -\frac{\mu_1}{2}\int_{\Sigma_2} q_{\Ba}\,\hat{\bm{\cP}}^{\Ba}_{1+1} 
 + \int_{\Sigma_2} \hat{\cL}^{\text{WZ}}_{(p,q)\text{-1}}
\nn\\
 &= \frac{1}{4}\int_{\Sigma_2} \Exp{\lambda}\cM_{IJ}\,\bm{\cP}^I\wedge *_\gamma\bm{\cP}^J 
 - \mu_1\int_{\Sigma_2}q_{\Ba}\,\bigl(\tfrac{1}{2}\,\bm{\cP}^{\Ba}_{m}\wedge Dx^m+F^{\Ba}_2\bigr) 
\nn\\
 &\quad + q_{\Ba} \int_{\Sigma_2} \bigl(\cA^{\gi} \wedge \tilde{\kk}^{\Ba(1)}_{\gi} +\tfrac{1}{2}\,\cA^{\gi_1\gi_2}\,\iota_{\gi_2}\tilde{\kk}^{\Ba(1)}_{\gi_1} \bigr) \,.
\end{align}
If we introduce two matrices
\begin{align}
 \eta_{IJ}\equiv 
 {\footnotesize\begin{pmatrix} 0 & q_{\Bb}\,\delta_{m}^{n} & 0 & \cdots \\
 q_{\Ba}\,\delta^{m}_{n} & 0 & 0 & \\
 0 & 0 & 0 \\ \vdots &&& \ddots 
 \end{pmatrix} }, \qquad
 \omega^{(F)}_{IJ}\equiv 
{\footnotesize\begin{pmatrix} 2\,q_{\Ba}\,F^{\Ba}_{mn} & -q_{\Bb}\,\delta_{m}^{n} & 0 & \cdots \\
 q_{\Ba}\,\delta^{m}_{n} & 0 & 0 & \\
 0 & 0 & 0 \\ \vdots &&& \ddots 
 \end{pmatrix} }, 
\end{align}
we can express the gauged action as
\begin{align}
 \hat{S}_{(p,q)\text{-1}} = \frac{1}{4}\int_{\Sigma_2}\Exp{\lambda}\cM_{IJ}\,\bm{\cP}^I\wedge *_\gamma\bm{\cP}^J 
 -\frac{\mu_1}{4}\int_{\Sigma_2} \omega^{(F)}_{IJ}\, P^I\wedge P^J 
 + \frac{\mu_1}{2}\int_{\Sigma_2} \eta_{IJ}\, \cA^I\wedge P^J \,.
\end{align}
This is a natural uplift of the gauged string action \eqref{eq:D-gBSM} to the $U$-duality setup. 
Further details of this gauged sigma model are studied later in section \ref{sec:string-detail}. 

\subsection{Gauged D3-brane action}

The Lagrangian for the ungauged D3-brane theory is given by
\begin{align}
 \cL_{(\text{D3})} = \tfrac{1}{8}\cM_{IJ} \,P^I\wedge *_{\gamma} P^J 
 - \tfrac{\mu_3}{4}\,\bigl[\hat{P}_{m_3}\wedge \rmd x^{m_3} 
 + \epsilon_{\Ba\Bb}\, \hat{P}_m^{\Ba}\wedge \rmd x^m\wedge \bigl(B_2^{\Bb}-F_2^{\Bb}\bigr)\bigr]
 + \cL^{\text{WZ}}_{(\text{D3})}\,,
\end{align}
where $\lambda$ is absorbed into $\gamma$ and
\begin{align}
 \cL^{\text{WZ}}_{(\text{D3})} 
 = \mu_3\,\bigl(B_4 -\tfrac{1}{2}\,\epsilon_{\Ba\Bb}\,B_2^{\Ba}\wedge F_2^{\Bb} - F_4\bigr)\,. 
\end{align}
Similar to the case of the M5-brane, if we ignore the WZ term, this can be extended to a gauge-invariant one as
\begin{align}
 \hat{\cL}_0 &= \tfrac{1}{8}\,\cM_{IJ} \,\bm{\cP}^I\wedge *_{\gamma} \bm{\cP}^J 
\nn\\
 &\quad - \tfrac{\mu_3}{4}\,\bigl[\hat{\bm{\cP}}_{1+3} 
 + \epsilon_{\Ba\Bb}\, \hat{\bm{\cP}}^{\Ba}_{1+1}\wedge \bigl(B_2^{\Bb}-F_2^{\Bb}+\cA^{\gi}\wedge \hat{\kk}_{\gi}^{\Bb(1)}+\tfrac{1}{2!}\,\cA^{\gi_1\gi_2}\,\iota_{\gi_2}\hat{\kk}_{\gi_1}^{\Bb(1)}\bigr)\bigr] \,.
\end{align}
By using
\begin{align}
 \delta B_4 = -\epsilon^{\gi}\,\bigl(\rmd \tilde{\kk}_{\gi}^{(3)} + \tfrac{1}{2}\,\epsilon_{\Bc\Bd}\,B^{\Bc}_2\wedge \rmd \tilde{\kk}_{\gi}^{\Bd(1)}\bigr) + \rmd \epsilon^{\gi}\wedge\iota_{\gi}B_4\,,
\end{align}
the variation of the WZ term can be found as
\begin{align}
 \mu_3^{-1}\,\delta \cL^{\text{WZ}}_{(\text{D3})} 
 &=\rmd\epsilon^{\gi}\wedge \bigl[\hat{\kk}_{\gi 3} 
 -\tfrac{1}{2}\,\epsilon_{\Ba\Bb}\,\bigl(B_2^{\Ba}-F_2^{\Ba}\bigr)\wedge \hat{\kk}^{\Bb(1)}_{\gi} \bigr]
\nn\\
 &-\rmd\bigl[\epsilon^{\gi}\, \bigl(\tilde{\kk}_{\gi}^{(3)} - \tfrac{1}{2}\,\epsilon_{\Ba\Bb}\, \tilde{\kk}^{\Ba(1)}_{\gi} \wedge F_2^{\Bb}\bigr)\bigr] - \delta F_4\,.
\end{align}
This suggests us to define the variation of $A_3$ as
\begin{align}
 \delta A_3 = \epsilon^{\gi}\, \bigl( \tfrac{1}{2}\,\epsilon_{\Ba\Bb}\, \tilde{\kk}^{\Ba(1)}_{\gi} \wedge F_2^{\Bb} -\tilde{\kk}_{\gi}^{(3)}\bigr)\,,
\end{align}
and then we obtain
\begin{align}
 \mu_3^{-1}\,\delta \cL^{\text{WZ}}_{(\text{D3})} 
 &=\rmd\epsilon^{\gi}\wedge \bigl[\hat{\kk}_{\gi 3} 
 -\tfrac{1}{2}\,\epsilon_{\Ba\Bb}\,(B_2^{\Ba}-F_2^{\Ba})\wedge \hat{\kk}^{\Bb(1)}_{\gi} \bigr]\,.
\end{align}
Similar to the M5-brane case, to compensate the first term, we consider adding a combination
\begin{align}
 \cL_1 \equiv \mu_3\,\bigl(\cA^{\gi}\wedge \hat{\kk}_{\gi}^{(3)}+\tfrac{1}{2!}\,\cA^{\gi_1\gi_2}\wedge \iota_{\gi_2}\hat{\kk}_{\gi_1}^{(3)}+\cdots +\tfrac{1}{4!}\,\cA^{\gi_1\cdots \gi_4}\wedge \iota_{\gi_4\cdots \gi_2} \hat{\kk}_{\gi_1}^{(3)}\bigr)\,,
\end{align}
and use the identities:
\begin{align}
 \delta \bigl(\cA^{\gi}\wedge \hat{\kk}_{\gi}^{(3)}\bigr) &= -\rmd\epsilon^{\gi}\wedge\hat{\kk}_{\gi}^{(3)} + \cA^{\gi_1}\wedge \rmd \epsilon^{\gi_2}\wedge\iota_{\gi_2}\hat{\kk}_{\gi_1}^{(3)}\,,
\nn\\
 \delta \bigl(\tfrac{1}{2!}\,\cA^{\gi_1\gi_2}\wedge \iota_{\gi_2}\hat{\kk}_{\gi_1}^{(3)}\bigr) 
 &= - \cA^{\gi_1}\wedge \rmd \epsilon^{\gi_2}\wedge\iota_{[\gi_2}\hat{\kk}_{\gi_1]}^{(3)}
 + \tfrac{1}{2!}\,\cA^{\gi_1\gi_2}\wedge\rmd \epsilon^{\gi_3}\wedge\iota_{\gi_3\gi_2}\hat{\kk}_{\gi_1}^{(3)} \,,
\nn\\
 \delta \bigl(\tfrac{1}{3!}\,\cA^{\gi_1\gi_2\gi_3}\wedge \iota_{\gi_3\gi_2}\hat{\kk}_{\gi_1}^{(3)}\bigr) &= 
 -\tfrac{1}{2!}\,\cA^{\gi_1\gi_2}\wedge \rmd \epsilon^{\gi_3}\wedge \iota_{[\gi_3\gi_2}\hat{\kk}_{\gi_1]}^{(3)}
 + \tfrac{1}{3!}\,\cA^{\gi_1\gi_2\gi_3}\wedge \rmd \epsilon^{\gi_4}\wedge\iota_{\gi_4\gi_3\gi_2}\hat{\kk}_{\gi_1}^{(3)}\,,
\nn\\
 \delta \bigl(\tfrac{1}{4!}\,\cA^{\gi_1\cdots \gi_4}\wedge \iota_{\gi_4\cdots \gi_2} \hat{\kk}_{\gi_1}^{(3)}\bigr) &= 
 -\tfrac{1}{3!}\,\cA^{\gi_1\gi_2\gi_3}\wedge \rmd \epsilon^{\gi_4}\wedge \iota_{[\gi_4\gi_3\gi_2}\hat{\kk}_{\gi_1]}^{(3)} \,.
\end{align}
To compensate the second term, we add
\begin{align}
 \cL_2= -\tfrac{\mu_3}{2}\,\epsilon_{\Ba\Bb}\,(B_2^{\Ba}-F_2^{\Ba})\wedge \bigl(\cA^{\gi}\wedge \hat{\kk}_{\gi}^{\Bb(1)}+\tfrac{1}{2!}\,\cA^{\gi_1\gi_2}\,\iota_{\gi_2}\hat{\kk}_{\gi_1}^{\Bb(1)} \bigr)\,.
\end{align}
whose variation is
\begin{align}
\begin{split}
 \mu_3^{-1}\,\delta \cL_2&= \rmd \epsilon^{\gi}\wedge \bigl[\tfrac{1}{2}\,\epsilon_{\Ba\Bb}\,(B_2^{\Ba}-F_2^{\Ba})\wedge \tilde{\kk}_{\gi}^{\Bb(1)}\bigr]
 +\cA^{\gi_1}\wedge \rmd \epsilon^{\gi_2}\wedge \bigl(-\tfrac{1}{2}\,\epsilon_{\Ba\Bb}\,\tilde{\kk}_{\gi_1}^{\Ba(1)}\wedge \hat{\kk}_{\gi_2}^{\Bb(1)}\bigr) 
\\
 &\quad +\cA^{\gi_1\gi_2}\wedge \rmd \epsilon^{\gi_3}\wedge \bigl(-\tfrac{1}{4}\,\epsilon_{\Ba\Bb}\,\tilde{\kk}_{\gi_3}^{\Ba(1)} \,\iota_{\gi_2}\hat{\kk}_{\gi_1}^{\Bb(1)} \bigr)\,.
\end{split}
\end{align}
Computing $\delta (\cL_1+\cL_2)$, we find that terms that are linear in $\cA^{\gi}$ are canceled out by using
\begin{align}
 \cA^{\gi_1}\wedge \rmd \epsilon^{\gi_2}\wedge\bigl(\iota_{(\gi_1}\hat{\kk}^{(3)}_{\gi_2)} 
 -\tfrac{1}{2}\,\epsilon_{\Ba\Bb}\,\hat{\kk}^{\Ba(1)}_{\gi_1}\wedge\hat{\kk}^{\Bb(1)}_{\gi_2}\bigr)=0\,,
\end{align}
which follows from the isotropy condition \eqref{eq:B3-SC}. 
To compensate the quadratic term, we need to add the term
\begin{align}
 \cL_3 \equiv \tfrac{\mu_3}{12} \,\cA^{\gi_1\gi_2\gi_3}\wedge \epsilon_{\Ba\Bb}\,\iota_{\gi_3}\hat{\kk}^{\Ba(1)}_{\gi_2}\,\hat{\kk}^{\Bb(1)}_{\gi_1}\,,
\end{align}
whose variation is
\begin{align}
\begin{split}
 \mu_3^{-1}\,\delta \cL_3 
 &= -\tfrac{1}{4}\,\cA^{\gi_1\gi_2}\wedge \rmd\epsilon^{\gi_3}\wedge \epsilon_{\Ba\Bb}\,\iota_{[\gi_3}\hat{\kk}^{\Ba(1)}_{\gi_2}\,\hat{\kk}^{\Bb(1)}_{\gi_1]}
\\
 &\quad + \tfrac{1}{12}\,\cA^{\gi_1\gi_2\gi_3}\wedge\rmd \epsilon^{\gi_4}\, \epsilon_{\Ba\Bb}\,\iota_{\gi_3}\hat{\kk}^{\Ba(1)}_{\gi_2}\,\iota_{\gi_1}\hat{\kk}^{\Bb(1)}_{\gi_4}\,.
\end{split}
\end{align}
The isotropy condition \eqref{eq:B3-SC} gives the identity
\begin{align}
 0 =\iota_{\gi_3\gi_2}\hat{\kk}^{(3)}_{\gi_1}-\iota_{\gi_2\gi_1}\hat{\kk}^{(3)}_{\gi_3}
 -\epsilon_{\Ba\Bb}\,\iota_{\gi_1}\hat{\kk}^{\Ba(1)}_{\gi_3}\,\hat{\kk}^{\Bb(1)}_{\gi_2} 
 +\epsilon_{\Ba\Bb}\,\hat{\kk}^{\Ba(1)}_{\gi_3}\,\iota_{\gi_1}\hat{\kk}^{\Bb(1)}_{\gi_2} \,, 
\end{align}
and then we find
\begin{align}
 0&=\tfrac{1}{3!}\,\cA^{\gi_1\gi_2}\wedge\rmd\epsilon^{\gi_3}\wedge \bigl(\iota_{\gi_3\gi_2}\hat{\kk}^{(3)}_{\gi_1}-\iota_{\gi_2\gi_1}\hat{\kk}^{(3)}_{\gi_3}
 -\epsilon_{\Ba\Bb}\,\iota_{\gi_1}\hat{\kk}^{\Ba(1)}_{\gi_3}\,\hat{\kk}^{\Bb(1)}_{\gi_2} 
 +\epsilon_{\Ba\Bb}\,\hat{\kk}^{\Ba(1)}_{\gi_3}\,\iota_{\gi_1}\hat{\kk}^{\Bb(1)}_{\gi_2}\bigr) 
\nn\\
  &=\tfrac{1}{3!}\,\cA^{\gi_1\gi_2}\wedge\rmd\epsilon^{\gi_3}\wedge \bigl(\iota_{\gi_3\gi_2}\hat{\kk}^{(3)}_{\gi_1}-\iota_{\gi_2\gi_1}\hat{\kk}^{(3)}_{\gi_3}\bigl)
\nn\\
  &\quad
 - \tfrac{1}{4}\,\cA^{\gi_1\gi_2}\wedge\rmd\epsilon^{\gi_3}\wedge \bigl(\epsilon_{\Ba\Bb}\,\hat{\kk}^{\Ba(1)}_{\gi_3}\,\iota_{\gi_2}\hat{\kk}^{\Bb(1)}_{\gi_1}
  + \epsilon_{\Ba\Bb}\,\iota_{[\gi_1}\hat{\kk}^{\Ba(1)}_{\gi_3}\,\hat{\kk}^{\Bb(1)}_{\gi_2]} \bigr) \,.
\end{align}
We can also show the identity
\begin{align}
 0&=\tfrac{1}{4!}\,\cA^{\gi_1\gi_2\gi_3}\wedge\rmd\epsilon^{\gi_4}\, \bigl(\iota_{\gi_4\gi_3\gi_2}\hat{\kk}^{(3)}_{\gi_1}+\iota_{\gi_3\gi_2\gi_1}\hat{\kk}^{(3)}_{\gi_4}
 +2\,\epsilon_{\Ba\Bb}\,\iota_{\gi_3}\hat{\kk}^{\Ba(1)}_{\gi_2}\,\iota_{\gi_1}\hat{\kk}^{\Bb(1)}_{\gi_4} \bigr) \,. 
\end{align}
Using these, we can check that the quadratic and the cubic terms are also canceled out, and we have shown the invariance of
\begin{align}
 \hat{\cL}^{\text{WZ}}_{(\text{D3})} 
 \equiv \cL^{\text{WZ}}_{(\text{D3})} + \cL_1 + \cL_2 + \cL_3\,.
\end{align}
Thus the Lagrangian for the gauged D3-brane theory is $\hat{\cL}_{(\text{D3})}\equiv \hat{\cL}_0+\hat{\cL}^{\text{WZ}}_{(\text{D3})}$\,. 

In summary, the gauged D3-brane action is given by
\begin{align}
 \hat{S}_{(\text{D3})}&= \tfrac{1}{8}\int_{\Sigma_4} \cM_{IJ} \,\bm{\cP}^I\wedge *_{\gamma} \bm{\cP}^J 
 +\mu_3\int_{\Sigma_4}\bigl(B_4 -\tfrac{1}{2}\,\epsilon_{\Ba\Bb}\,B_2^{\Ba}\wedge F_2^{\Bb} - F_4\bigr)
\nn\\
 &\quad + \mu_3\int_{\Sigma_4}\bigl(\cA^{\gi}\wedge \hat{\kk}_{\gi}^{(3)}+\tfrac{1}{2!}\,\cA^{\gi_1\gi_2}\wedge \iota_{\gi_2}\hat{\kk}_{\gi_1}^{(3)}+\cdots +\tfrac{1}{4!}\,\cA^{\gi_1\cdots \gi_4}\wedge \iota_{\gi_4\cdots \gi_2} \hat{\kk}_{\gi_1}^{(3)} -\tfrac{1}{4}\,\hat{\bm{\cP}}_{1+3} \bigr)
\nn\\
 &\quad -\tfrac{\mu_3}{2}\int_{\Sigma_4}\epsilon_{\Ba\Bb}\,\bigl(B_2^{\Ba}-F_2^{\Ba}+\tfrac{1}{2}\,\hat{\bm{\cP}}^{\Ba}_{1+1}\bigr)\wedge \bigl(\cA^{\gi}\wedge \hat{\kk}_{\gi}^{\Bb(1)}+\tfrac{1}{2!}\,\cA^{\gi_1\gi_2}\,\iota_{\gi_2}\hat{\kk}_{\gi_1}^{\Bb(1)} -\tfrac{1}{2}\,\hat{\bm{\cP}}^{\Bb}_{1+1}\bigr)
\nn\\
 &\quad + \tfrac{\mu_3}{12}\int_{\Sigma_4}\cA^{\gi_1\gi_2\gi_3}\wedge \epsilon_{\Ba\Bb}\,\iota_{\gi_3}\hat{\kk}^{\Ba(1)}_{\gi_2}\,\hat{\kk}^{\Bb(1)}_{\gi_1}\,.
\end{align}
Here again, we have repeatedly used the isotropy conditions, \eqref{eq:B1-SC} and \eqref{eq:B3-SC}. 
The structure is very similar to the gauged M5-brane action, but due to the dimensionality, this is slightly simpler than the M5-brane action. 

We hope that a similar analysis can be done for a $(p,q)$-5-brane, but this will be much more involved and we shall leave it to future work. 

\section{Reduced background fields}
\label{sec:RBF}

In this section, we study the supergravity fields on the exceptional dressing cosets. 

\subsection{Standard $(p,q)$-string action}
\label{sec:string-detail}

In the previous section, we have obtained the gauged string action as
\begin{align}
 \hat{S}_{(p,q)\text{-1}}&= \frac{1}{4}\int_{\Sigma_2} \Exp{\lambda}\cM_{IJ}\,\bm{\cP}^I\wedge *_\gamma\bm{\cP}^J 
 -\frac{\mu_1}{2}\int_{\Sigma_2} q_{\Ba}\,\hat{\bm{\cP}}^{\Ba}_{m}\wedge Dx^m 
\nn\\
 &\quad + \mu_1\int_{\Sigma_2} q_{\Ba}\,\bigl(B^{\Ba}_2-F^{\Ba}_2+\cA^{\gi}\wedge \hat{\kk}_{\gi}^{\Ba(1)}+\tfrac{1}{2!}\,\cA^{\gi_1\gi_2}\,\iota_{\gi_2}\hat{\kk}_{\gi_1}^{\Ba(1)}\bigr)\,.
\end{align}
By using the equations of motion for the auxiliary fields $\hat{\bm{\cP}}_{m_3}$, $\hat{\bm{\cP}}_{m_5}^{\Ba}$, $\cdots$\,, this reduces to
\begin{align}
 S&= \frac{1}{4}\int_{\Sigma_2} \bigl(\Exp{\lambda}\mathsf{g}_{mn}\,D x^m\wedge *_\gamma D x^n + \Exp{\lambda}m_{\Ba\Bb}\,\mathsf{g}^{mn}\,\widehat{\bm{\cP}}^{\Ba}_m\wedge *_\gamma\widehat{\bm{\cP}}^{\Bb}_n -2\,\mu_1\,q_{\Ba}\,\hat{\bm{\cP}}^{\Ba}_{m} \wedge D x^m\bigr)
\nn\\
 &\quad + \mu_1\int_{\Sigma_2} q_{\Ba}\,\bigl(B^{\Ba}_2-F^{\Ba}_2+\cA^{\gi}\wedge \hat{\kk}_{\gi}^{\Ba(1)}+\tfrac{1}{2!}\,\cA^{\gi_1\gi_2}\,\iota_{\gi_2}\hat{\kk}_{\gi_1}^{\Ba(1)}\bigr)\,.
\end{align}
The equations of motion for $\hat{\bm{\cP}}^{\Ba}_m$ give
\begin{align}
 \widehat{\bm{\cP}}^{\Ba}_m = \mu_1\Exp{-\lambda}m^{\Ba\Bb}\,q_{\Bb}\,\mathsf{g}_{mn}\,*_\gamma D x^n \,.
\end{align}
Eliminating $\widehat{\bm{\cP}}^{\Ba}_m$\,, we obtain
\begin{align}
 S&= \frac{1}{4}\int_{\Sigma_2} \bigl(\Exp{\lambda} + \Exp{-\lambda}\mu_1^2\, q_{\Ba}\,m^{\Ba\Bb}\,q_{\Bb} \bigr)\, \mathsf{g}_{mn}\,Dx^{m}\wedge *_\gamma Dx^n 
\nn\\
 &\quad + \mu_1\int_{\Sigma_2} q_{\Ba}\,\bigl(B^{\Ba}_2-F^{\Ba}_2+\cA^{\gi}\wedge \hat{\kk}_{\gi}^{\Ba(1)}+\tfrac{1}{2!}\,\cA^{\gi_1\gi_2}\,\iota_{\gi_2}\hat{\kk}_{\gi_1}^{\Ba(1)}\bigr)\,.
\end{align}
The equations of motion for $\lambda$ become
\begin{align}
 \Exp{\lambda} = \abs{\mu_1}\,\abs{q}\equiv \abs{\mu_1}\,\sqrt{q_{\Ba}\,m^{\Ba\Bb}\,q_{\Bb}}\,,
\end{align}
and then the action reduces to
\begin{align}
 S&= \frac{\abs{\mu_1}}{2}\int_{\Sigma_2} \abs{q}\, \mathsf{g}_{mn}\,Dx^{m}\wedge *_\gamma Dx^n + \mu_1\int_{\Sigma_2}q_{\Ba}\,\bigl(B^{\Ba}_2-F^{\Ba}_2\bigr)
\nn\\
 &\quad + \mu_1\int_{\Sigma_2} q_{\Ba}\,\bigl(\cA^{\gi}\wedge \hat{\kk}_{\gi}^{\Ba(1)}+\tfrac{1}{2!}\,\cA^{\gi_1\gi_2}\,\iota_{\gi_2}\hat{\kk}_{\gi_1}^{\Ba(1)}\bigr)\,.
\end{align}
If we consider the specific case $q_{\Ba}=(1,0)$, this reproduces the action by Hull and Spence \cite{Hull:1989jk}. 

For a general $q_{\Ba}$, by eliminating the gauge fields $\cA^{\gi}$, the gauged action reduces to the standard action
\begin{align}
 S = \frac{\abs{\mu_1}}{2}\int_{\Sigma_2} \abs{\check{q}}\, \check{\mathsf{g}}_{mn}\,\rmd x^{m}\wedge *_\gamma \rmd x^n + \mu_1\int_{\Sigma_2}q_{\Ba}\,\bigl(\check{B}^{\Ba}_2-F^{\Ba}_2\bigr) \,,
\end{align}
defined on an exceptional dressing coset with the background fields $\{\check{\mathsf{g}}_{mn},\,\check{B}^{\Ba}_2,\,\check{m}_{\Ba\Bb}\}$\,. 
Using the result of \cite{2112.14766}, we can express $\{\check{\mathsf{g}}_{mn},\,\check{B}^{\Ba}_2,\,\check{m}_{\Ba\Bb}\}$ in terms of the original background fields $\{\mathsf{g}_{mn},\,B^{\Ba}_2,\,m_{\Ba\Bb}\}$ and the generalized Killing vector fields $\KK_{\gi}{}^I$. 
For example, if a matrix
\begin{align}
 N_{\gi\gj} \equiv \abs{q}\,\kk_{\gi}^m\,\mathsf{g}_{mn}\,\kk_{\gj}^n + \iota_{\gj}\hat{\kk}^{\Ba}_{\gi}\,q_{\Ba}\,,
\end{align}
is invertible, by defining
\begin{align}
 E_{mn}\equiv \abs{q}\, \mathsf{g}_{mn} + q_{\Ba}\,B^{\Ba}_2\,, 
\end{align}
we can express the reduced background fields as
\begin{align}
 \check{E}^{(q)}_{mn} = E_{mn} -\bigl(\kk_{\gi m}-q_{\Ba}\,\hat{\kk}^{\Ba}_{\gi m}\bigr)\,N^{\gi\gj}\,\bigl(\kk_{\gj n}+q_{\Bb}\,\hat{\kk}^{\Bb}_{\gj n}\bigr)\,,
\label{eq:check-E}
\end{align}
where $\kk_{\gi m}\equiv \abs{\check{q}}\, \check{\mathsf{g}}_{mn}\,\kk_{\gi}^n$\,. 
Even when $N_{\gi\gj}$ is not invertible, we can still find an expression for $\check{E}_{mn}$ by means of the original fields $E_{mn}$ and the generalized Killing vector fields $\KK_{\gi}{}^I$ \cite{2112.14766}.

\subsection{An issue}

Here we note that the formula \eqref{eq:check-E} raises a puzzling issue: each gauged brane action may give different background fields on the exceptional dressing coset. 

For simplicity, let us consider the case where the original background fields $\{\mathsf{g}_{mn},\,B^{\Ba}_2,\,m_{\Ba\Bb}\}$, which are constructed from the EDA, satisfy $B^{\bm{2}}_2=0$ and $m_{\bm{1}\bm{2}}=0$ ($\Leftrightarrow\, C_0=C_2=0$) (i.e., only the metric, the $B$-field, and the dilaton are non-vanishing). 
We also suppose that the generalized Killing vector fields satisfy $\hat{\kk}^{\bm{2}}_{\gi m}=0$\,. 
Then the situation studied in standard dressing coset (associated with a Drinfel'd double) is realized. 
In this case, if we consider an F-string $q_{\Ba}=(1,0)$, the formula \eqref{eq:check-E} reproduces the result of \cite{2112.14766}, which determines the string-frame metric $g_{mn}\equiv \Exp{\frac{\Phi}{2}}\mathsf{g}_{mn}$ and the $B$-field on the dressing coset. 
Explicitly, the string-frame metric on the dressing coset becomes
\begin{align}
 \check{g}^{(1,0)}_{mn} = g_{mn} -\bigl(\kk_{\gi (m|}-\hat{\kk}^{\bm{1}}_{\gi (m|}\bigr)\,N^{\gi\gj}\,\bigl(\kk_{\gj |n)}+ \hat{\kk}^{\bm{1}}_{\gj |n)}\bigr) \qquad \bigl(\kk_{\gi m}\equiv g_{mn}\,\kk_{\gi}^n\bigr)\,,
\label{eq:F1-check-g}
\end{align}
where $N_{\gi\gj} = \kk_{\gi}^m\,g_{mn}\,\kk_{\gj}^n + \iota_{\gj}\hat{\kk}^{\bm{1}}_{\gi}$\,. 
Instead, if we consider a D-string $q_{\Ba}=(0,1)$ on the same original background, after eliminating the gauge fields $\cA_{\gi}$, we obtain the D-string action on a background with the metric given by the formula \eqref{eq:check-E}. 
Under our assumptions, we have $E_{mn}= \Exp{- \Phi}\, g_{mn}$ and the formula \eqref{eq:check-E} reduces to
\begin{align}
 \Exp{-\check{\Phi}^{(0,1)}}\, \check{g}^{(0,1)}_{mn}
 = \Exp{- \Phi}\, \bigl(g_{mn} - g_{mp}\,\kk_{\gi}^p \,n^{\gi\gj}\, \kk_{\gj}^q\,g_{qn}\bigr) 
 \qquad \bigl(n_{\gi\gj} \equiv \kk_{\gi}^m\,g_{mn}\,\kk_{\gj}^n\bigr)\,.
\label{eq:D1-check-g}
\end{align}
Namely, the reduced metric $\check{g}^{(1,0)}_{mn}$ given in Eq.~\eqref{eq:F1-check-g} that is observed by the F-string is different from $\check{g}^{(0,1)}_{mn}$ given in Eq.~\eqref{eq:D1-check-g} that is observed by the D-string. 

By the construction, each of the reduced metrics associated with the F-string or the D-string can be regarded as a metric on an exceptional dressing coset. 
However, the non-uniqueness (or the $q_{\Ba}$-dependence) of the reduced background fields $\check{E}^{(q)}_{mn}$ makes it difficult to discuss the generalized $U$-duality as a symmetry of the supergravity equations of motion. 
For a given EDA, we can decompose the EDA into a pair of algebras $\mathfrak{d}(\mathfrak{e}_{\nM(\nM)})=\mathfrak{g}\oplus\tilde{\mathfrak{g}}$ and construct the supergravity fields on the physical space $G$ by following the standard procedure. 
We then consider the gauged F-string action and obtain the metric and the $B$-field on the exceptional dressing coset $F\backslash G$. 
We then change the decomposition as $\mathfrak{d}(\mathfrak{e}_{\nM(\nM)})=\mathfrak{g}'\oplus\tilde{\mathfrak{g}}'$ and similarly obtain the metric and the $B$-field on another dressing coset $F\backslash G'$. 
To discuss the generalized $U$-duality at the level of supergravity equations of motion, we need to know all of the supergravity fields on these dressing cosets, but as long as we are considering the bosonic action, the F-string couples only to the metric and the $B$-field. 
To determine other fields, one may consider using other branes as additional probes, but as we have observed above, each brane perceives different supergravity fields and we cannot determine the set of supergravity fields on a single background.
To discuss the generalized $U$-duality at the level of supergravity, it may be more useful to take the following completely different approach. 

\subsection{Generalized $U$-duality of exceptional dressing cosets}
\label{sec:example}

In the study of the PL $T$-duality of dressing cosets, a useful procedure to determine the supergravity fields has been proposed by Sfetsos \cite{hep-th:9904188}. 
Recently, it was found in \cite{1903.00439,2112.14766} that this procedure always gives the same result as the approach based on the gauged sigma model.
Here we propose to apply this procedure to the exceptional dressing cosets. 

In the approach by Sfetsos, we assume that the gauge group $F$ is initially a subgroup of the Lie group $G$, which corresponds to the physical space $M$. 
Namely, we start with a standard coset $F\backslash G$.
After the generalized duality transformation, this may be mapped to a certain dressing coset $F\backslash G'$. 
The procedure to construct the supergravity fields is very simple, and in the following, we shall explain the procedure by extending this to the $U$-duality setup. 

As in the case of the usual generalized $U$-duality, we construct the generalized metric as $\cM_{IJ}\propto E_I{}^A\,E_J{}^B\,\hat{\cM}_{AB}$ by using the generalized frame fields $E_A{}^I(x)$ and a constant matrix $\hat{\cM}_{AB}\in E_{\nM(\nM)}$. 
The difference from the usual procedure is that we introduce a real parameter $\lambda$ into the matrix $\hat{\cM}_{AB}$ such that the metric on the group manifold $G$ reduces to that of the coset space $F\backslash G$. 
For simplicity, here we shall consider a purely gravitational configuration, and then the metric is constructed as $\mathsf{g}_{mn}=e_m^a\,e_n^b\,\hat{\mathsf{g}}_{ab}$, where $\hat{\mathsf{g}}_{ab}$ is a constant matrix that construct the matrix $\hat{\cM}_{AB}$. 
We decompose the generators $T_a$ of $\mathfrak{g}$ into the generators $T_{\gi}$ of $\mathfrak{f}$ and the other generators $T_{r}$ ($r=1,\dotsc,D-n$), and introduce the parameter $\lambda$ as
\begin{align}
 \hat{\mathsf{g}}_{ab} = \begin{pmatrix} \hat{\mathsf{g}}_{rs} & 0 \\ 0 & \hat{\mathsf{g}}_{\gi\gj}\end{pmatrix}
 \quad \to\quad \begin{pmatrix} \hat{\mathsf{g}}_{rs} & 0 \\ 0 & \lambda\,\hat{\mathsf{g}}_{\gi\gj}
\end{pmatrix}. 
\end{align}
where we have assumed $\hat{\mathsf{g}}_{r\gj}=0$ just for simplicity. 
If we take the Sfetsos limit $\lambda\to 0$, the metric becomes degenerate and we obtain the metric $\mathsf{g}_{mn}=e_m^{r}\,e_n^s\,\hat{\mathsf{g}}_{rs}$ on the coset space $F\backslash G$. 
Before taking the limit, we perform a generalized $U$-duality and obtain the supergravity fields on the dual geometry.
We then take limit $\lambda\to 0$ and also make a certain gauge fixing, and then obtain the supergravity fields on the exceptional dressing coset $F\backslash G'$. 

To make it more explicit, let us use a familiar example of non-Abelian $T$-duality of 2-sphere $\text{S}^2\simeq \UU(1)\backslash\SO(3)$ \cite{hep-th:9210021}. 
We begin with the type IIB EDA $\mathfrak{d}(\mathfrak{e}_{\nM(\nM)})$ ($5\leq \nM\leq 8$) (see section 6 of \cite{2009.04454} for the details) with the only non-vanishing structure constants of $\mathfrak{so}(3)$
\begin{align}
 f_{12}{}^3 = 1\,,\quad
 f_{23}{}^1 = 1\,,\quad
 f_{13}{}^2 = -1\,.
\end{align}
Introducing the local coordinates as
\begin{align}
 (x^m)=(\theta,\,\phi,\,\psi,\,y^1,\dotsc,\,y^{\nM-4})\,,
\end{align}
we parameterize the group element of $G$ as
\begin{align}
 g=\Exp{\psi\,T_3}\Exp{\theta\,T_1}\Exp{-\phi\,T_3} \Exp{y^1\,T_4+\cdots+y^{\nM-4}\,T_{\nM-1}}.
\end{align}
Then we obtain the generalized frame fields of the form
\begin{align}
 E_A{}^I = \begin{pmatrix} e_a{}^m & 0 &0 \\
 0 & \delta_{\Ba}^{\Bb}\,r^a{}_m & 0 \\ 0 & 0 & \ddots\end{pmatrix},
\end{align}
where
\begin{align}
 e_a{}^m = \begin{pmatrix} \cos \psi & -\frac{\sin \psi}{\sin\theta} & - \frac{\sin \psi}{\tan\theta} & \bm{0} \\
 \sin \psi & \frac{\cos \psi}{\sin\theta} & \frac{\cos \psi}{\tan \theta} &\bm{0} \\ 0 & 0 & 1 & \bm{0} \\ \bm{0} & \bm{0} & \bm{0} &\bm{1}_{\nM-4} & \end{pmatrix}.
\end{align}
We introduce a constant metric
\begin{align}
 \hat{\mathsf{g}}_{ab} = \diag(1,1,\lambda,1,\dotsc,1)\,,
\end{align}
and constant the matrix $\hat{\cM}_{AB}$ as
\begin{align}
 \hat{\cM}_{AB}= \abs{\hat{\mathsf{g}}}^{\frac{1}{9-n}}\begin{pmatrix} \hat{\mathsf{g}}_{ab} & 0 & 0 & \\
 0 & \delta_{\Ba\Bb}\,\hat{\mathsf{g}}^{ab} & 0 &\cdots \\
 0 & 0 & \hat{\mathsf{g}}^{a_3b_3} & \\
 & \vdots & & \ddots 
\end{pmatrix}.
\label{eq:hat-M-original}
\end{align}
The Sfetsos limit is defined by $\lambda\to 0$ but we do not take the limit at this stage. 
We construct the generalized metric as $\cM_{IJ}\propto E_I{}^A\,E_J{}^B\,\hat{\cM}_{AB}$, where the overall factor is determined such that $\cM_{IJ}$ has a unit determinant. 
Then, we can easily see that only the (Einstein-frame) metric $\mathsf{g}_{mn}=r_m^a\,r_n^b\,\hat{\mathsf{g}}_{ab}$ is non-vanishing. 
The explicit form of the metric is
\begin{align}
 \rmd \mathsf{s}^2 = \rmd\theta^2+\sin^2\theta\,\rmd\phi^2 + \lambda\,(\rmd\psi-\cos\theta\,\rmd\phi)^2 + (\rmd y^1)^2 +\cdots+(\rmd y^{\nM-4})^2\,.
\label{eq:metric-SO3}
\end{align}
This background admits the generalized Killing vector field
\begin{align}
 \KK \equiv \KK_{\gi=1} = E_3 = \partial_\psi\,,
\end{align}
and this $\UU(1)$ isometry direction is gauged. 
This gauge symmetry can be fixed as $\psi=0$. 
In this case, both of the formulas for the F-string \eqref{eq:F1-check-g} and the D-string \eqref{eq:D1-check-g} give the same reduced metric on 2-sphere (with Abelian directions) as
\begin{align}
 \rmd \check{\mathsf{s}}^2 = \rmd\theta^2+\sin^2\theta\,\rmd\phi^2 + (\rmd y^1)^2 +\cdots+(\rmd y^{\nM-4})^2\,.
\end{align}
The same result can be obtained also by taking the Sfetsos limit $\lambda\to 0$ in Eq.~\eqref{eq:metric-SO3}. 

Now, we perform the factorized $T$-dualities along $T_1$-, $T_2$-, $T_3$-, and $T_4$-directions, and obtain a new type IIB EDA with the non-vanishing structure constants
\begin{align}
 f^{\bm{1}}_3{}^{12} = -1\,,\qquad
 f^{\bm{1}}_1{}^{23} = -1\,,\qquad
 f^{\bm{1}}_2{}^{13} = 1\,.
\end{align}
Using the coordinates $(x^m)=(\rho,\,\theta,\,z,\,y^1,\dotsc,\,y^{\nM-4})$, we parameterize the group element as
\begin{align}
 g= \Exp{z\,T_3}\Exp{\rho\sin\theta\,T_2}\Exp{\rho\cos\theta\,T_1} \Exp{y^1\,T_4+\cdots +y^{\nM-4}\,T_{\nM-1}} .
\end{align}
We can again systematically construct the generalized frame fields $E'_A{}^I$ and construct the generalized metric as
\begin{align}
 \cM'_{IJ} \propto E'_I{}^A\,E'_J{}^B\,\hat{\cM}'_{AB}\,.
\end{align}
Here, $\hat{\cM}'_{AB}$ can be obtained by acting the factorized $T$-dualities to the original $\hat{\cM}_{AB}$ given in Eq.~\eqref{eq:hat-M-original}. 
We rewrite this generalized metric into the standard form $\cM_{IJ}=(L^{\rmT}\,\hat{G}\,L)_{IJ}$ (where we have removed the primes) and find that the matrix $L^I{}_J$ contains the $B$-field
\begin{align}
 B^{\bm{1}}_2 =\frac{-\lambda\,z\,\rho\,\rmd\rho\wedge\rmd \theta - \rho^2\,\rmd \theta\wedge\rmd z}{\rho^2 +\lambda\,(1+z^2)}\,.
\end{align}
The block-diagonal matrix $\hat{G}$ can be parameterized as
\begin{align}
 \hat{G}_{IJ} &= \abs{\mathsf{g}}^{\frac{1}{9-n}}\begin{pmatrix}
 \mathsf{g}_{mn} & 0 & 0 & \cdots \\
 0 & m_{\Ba\Bb}\,\mathsf{g}^{mn} & 0 & \cdots \\
 0 & 0 & \mathsf{g}^{m_3n_3} & \cdots \\
 \vdots & \vdots & \vdots & \ddots
\end{pmatrix},
\\
 m_{\Ba\Bb} &=
\begin{pmatrix}
 1 & -C_0 \\ 0 & 1
\end{pmatrix}
\begin{pmatrix}
 \Exp{-\Phi} & 0 \\ 0 & \Exp{\Phi}
\end{pmatrix}
\begin{pmatrix}
 1 & 0 \\ -C_0 & 1
\end{pmatrix},
\end{align}
and we find the Einstein-frame metric and the dilaton as
\begin{align}
\begin{split}
 \mathsf{g}_{mn}&= \Exp{-\frac{\Phi}{2}}\begin{pmatrix}
 \frac{\rho^2+\lambda}{\rho^2+\lambda\,(1+z^2)} & 0 & \frac{\rho\,z}{\rho^2+\lambda\,(1+z^2)} & 0 & \cdots & 0 \\
 0 & \frac{\lambda\,\rho^2}{\rho^2+\lambda\,(1+z^2)} & 0 & 0 & \cdots & 0 \\
 \frac{\rho\,z}{\rho^2+\lambda\,(1+z^2)} & 0 & \frac{1+z^2}{\rho^2+\lambda\,(1+z^2)} & 0 & \cdots & 0 \\
 0 & 0 & 0 & 1 & \cdots & 0 \\
 \vdots & \vdots & \vdots & \vdots & \ddots & 0 \\
 0 & 0 & 0 & 0 & 0 & 1 \end{pmatrix},
\\
 \Exp{-2\Phi}&= \rho^2+\lambda\,(1+z^2)\,.
\end{split}
\label{eq:dual-S2}
\end{align}
In this duality frame, the generalized Killing vector field $\KK_{\gi=1}$ that generates the dressing action becomes
\begin{align}
 \KK = E^3_{\bm{1}} = \partial_\theta + \tilde{\kk}^{\Ba}_m\,\partial^m_{\Ba} \quad\text{ with }\quad \tilde{\kk}^{\bm{1}}_m\,\rmd x^m =\rmd z\,,\quad  \tilde{\kk}^{\bm{2}}_m\,\rmd x^m =0\,,
\end{align}
and by using the gauge symmetry, we can impose the gauge-fixing condition as $\theta=0$\,. 
By taking the Sfetsos limit $\lambda\to 0$\,, we obtain the supergravity fields as
\begin{align}
\begin{split}
 \rmd s^2 &= \rmd\rho^2 + \frac{2z\,\rmd\rho\,\rmd z}{\rho} + \frac{1+z^2}{\rho^2}\,\rmd z^2 + (\rmd y^1)^2 + \cdots + (\rmd y^{\nM-4})^2\,,
\\
 \Exp{-2\Phi} &= \rho^2\,,\qquad B_2=0\,,
\end{split}
\label{eq:dual-S2-limit}
\end{align}
where the metric corresponds to the string-frame metric. 
This is exactly the same result as that of non-Abelian $T$-duality, but the treatment of the dilaton is different. 
In non-Abelian $T$-duality, the metric and the $B$-field are determined from the gauged action, but the dilaton is determined differently. 
However, in ExFT, all of the supergravity fields are contained in the generalized metric $\cM_{IJ}$ and can be determined in a uniform manner. 

We note that the background \eqref{eq:dual-S2} can be reproduced from the gauged F1-action, but the gauged D1-action gives a different metric for $\lambda\neq 0$. 
Only if we take the Sfetsos limit $\lambda\to 0$, the consistent result can be obtained from the gauged D1-action, but in this case, the gauge fields $\cA^{\gi}$ completely disappear from the gauged D1-action and the consistency is trivial. 
At the present stage, the consistency between the gauged action and the Sfetsos limit is still unclear and needs to be clarified in future study. 

At least when we discuss the supergravity equations of motion, the approach based on the Sfetsos limit would be useful. 
Once we identify how to introduce $\lambda$ in the original duality frame, the dualization procedure is the same as the usual generalized $U$-duality. 
After the dualization, we only need to make a gauge fixing and take the Sfetsos limit $\lambda\to 0$. 
A possible subtlety is that, in order to make \eqref{eq:dual-S2-limit} a ten-dimensional metric, we need to consider $\nM=12$, which goes beyond $E_{11}$. 
Of course, this is not a problem.
Rather than considering such an extension, we can introduce the external space that is not considered in this paper. 

Here we have only reproduced the known result, but it would not be difficult to find new examples of the generalized $U$-duality of dressing cosets. 
We begin with a standard coset space satisfying the supergravity equations of motion. 
After the dualization, the generalized Killing vector fields $\KK_{\gi}{}^I$ are mapped to new generalized Killing vector fields that generate the dressing action in the dual geometry. 
The Sfetsos limit will correspond to gauging the dressing action, and we obtain the exceptional dressing coset. 
A natural question is whether the dual geometry satisfies the supergravity equations of motion. 
This has not yet been answered completely even at the level of the generalized $U$-duality of group manifolds. 
We hope to address this question in the future by using the flux-formulation of ExFT or a similar formulation.

\section{Conclusions}
\label{sec:conclusions}

In this paper, we have proposed a definition of the dressing cosets associated with the EDA, which we have called the exceptional dressing coset. 
The standard definition, as a double coset $F\backslash \cD/\tilde{G}$, cannot be applied to EDA because the correspondent of the Lie group $\cD$ has not been defined for the EDA. 
Instead of considering a double coset, we have defined the exceptional dressing coset by gauging the dressing action on the physical space $G$. 
In the case of the Drinfel'd double, the dressing action is generated by the generalized Lie derivative $\epsilon^{\gi} \gLie_{\KK_{\gi}}$ in DFT along some generalized Killing vector field $\epsilon^{\gi}\,\KK_{\gi}$. 
As a natural extension, we have defined the $U$-duality version of the dressing action as the transformation that is generated by the generalized Lie derivative $\epsilon^{\gi} \gLie_{\KK_{\gi}}$ in ExFT. 

To see if this makes sense, it will be important to construct the gauged sigma models for various (single) branes in M-theory and type IIB theory. 
At least for the membrane, the M5-brane, the $(p,q)$-string, and the D3-brane, we have found that the gauged sigma model can be defined in an arbitrary curved $D$-dimensional background $M$ admitting $n$ generalized vector fields $\KK_{\gi}{}^I$ that (i) satisfy the generalized Killing equation $\gLie_{\KK_{\gi}}\cM_{IJ}=0$, (ii) are orthogonal to each other with respect to the bilinear forms which define the section condition of ExFT, and (iii) form a closed algebra by utilizing the generalized Lie derivative $\gLie_{\KK_{\gi}}\KK_{\gj}{}^I=-f_{\gi\gj}{}^{\gk}\,\KK_{\gk}{}^I$. 
In particular, if we consider a $D$-dimensional group manifold $M\simeq G$ that is constructed from the EDA, we can naturally construct the generalized vector fields $\KK_{\gi}{}^I$ which generate the dressing action and satisfy the above conditions (i)--(iii). 
Then, eliminating the gauge fields $\cA^{\gi}$ (by using the equations of motion), we can in principle obtain the standard brane actions whose target space is a $(D-n)$-dimensional exceptional dressing coset. 

For the gauged $(p,q)$-string action, we have eliminated the gauge fields $\cA^{\gi}$ and obtained the standard worldsheet action of a $(p,q)$-string on an exceptional dressing coset. 
However, from this action, we can determine only the supergravity fields that couple to the $(p,q)$-string. 
In addition, we have found that the supergravity fields computed from the F-string action and that from the D-string action can be generally different.
This makes it difficult to discuss the generalized $U$-duality at the level of the supergravity equations of motion. 
To discuss the generalized $U$-duality, we found it more reasonable to consider an extension of Sfetsos's approach to the $U$-duality setup. 
This allows us to compute all of the bosonic fields on the exceptional dressing cosets, and we can, in principle, check whether a solution of ExFT is mapped to another solution under the generalized $U$-duality transformation. 
As a demonstration, we have reproduced the well-known results of non-Abelian $T$-duality of 2-sphere in the $U$-duality-covariant formalism. 
Unlike the case of the $T$-duality, we found that the dilaton field is also determined in the same manner as the metric and the $B$-field. 

In this paper, we have made two proposals on the exceptional dressing cosets: one is based on the gauged brane actions and the other is based on the Sfetsos limit. 
In the case of the PL $T$-duality, these two approaches give the same result, but in the $U$-duality case, they do not always give the same results. 
This subtlety needs to be clarified. 
In future work, it is interesting to investigate non-trivial examples of the generalized $U$-duality of exceptional dressing cosets. 
It is also important to show whether the generalized $U$-duality of the exceptional dressing coset is a solution generating transformation in ExFT. 

\subsection*{Acknowledgments}

This work is supported by JSPS Grant-in-Aids for Scientific Research (C) 18K13540 and (B) 18H01214. 

\newpage

\appendix

\section{Notations}
\label{app:notation}

In this paper (after section \ref{sec:review-dressing-cosets}), we have used the multiple-index notation introduced in \cite{2004.09486}. 
When we denote $A_{\sfi_p}$, this corresponds to $\frac{A_{\sfi_1\cdots \sfi_p}}{\sqrt{p!}}$ in the standard notation. 
The antisymmetrized Kronecker delta includes additional factorial and $\delta_{\sfi_p}^{\sfj_p}$ reads $\frac{p!\,\delta_{\sfi_1\cdots \sfi_p}^{\sfj_1\cdots \sfj_p}}{\sqrt{p!\,p!}}=\delta_{\sfi_1\cdots \sfi_p}^{\sfj_1\cdots \sfj_p}$\,, where $\delta_{\sfi_1\cdots \sfi_p}^{\sfj_1\cdots \sfj_p}\equiv \frac{1}{p!}\,(\delta_{\sfi_1}^{\sfj_1}\cdots \delta_{\sfi_p}^{\sfj_p}\pm\,\text{permutations})$. 
For curved indices $\sfi$, which correspond to $i$ (in M-theory) and $m$ (in type IIB theory), we define the square bracket as
\begin{align}
 A_{\sfi_p[\sfj_q}\,B_{\sfj_r]} \equiv \delta^{\sfk_{q}\sfl_r}_{\sfj_{q+r}}\,A_{\sfi_p \sfk_{q}}\,B_{\sfl_r}\,.
\end{align}
Here, the right-hand side reads
\begin{align}
 \frac{(q+r)!\,\delta^{\sfk_1\cdots \sfk_q\sfl_1\cdots \sfl_r}_{\sfj_1\cdots \sfj_{q+r}}\,A_{\sfi_1\cdots \sfi_p \sfk_1\cdots \sfk_{q}}\,B_{\sfl_1\cdots \sfl_r}}{\sqrt{q!\,r!\,(q+r)!\,p!\,q!\,r!}}
 =\frac{\frac{(q+r)!}{q!\,r!}\,\delta^{\sfk_1\cdots \sfk_q\sfl_1\cdots \sfl_r}_{\sfj_1\cdots \sfj_{q+r}}\,A_{\sfi_1\cdots \sfi_p \sfk_1\cdots \sfk_{q}}\,B_{\sfl_1\cdots \sfl_r}}{\sqrt{(q+r)!\,p!}}\,.
\end{align}
We also use a condensed notations, such as
\begin{align}
 \rmd x^{\sfi_p} \to \frac{1}{\sqrt{p!}}\,\rmd x^{\sfi_1}\wedge\cdots \rmd x^{\sfi_p}\,,\qquad
 Dx^{\sfi_p} \to \frac{1}{\sqrt{p!}}\,D x^{\sfi_1}\wedge\cdots D x^{\sfi_p}\,. 
\end{align}
Then, for example, $A_{\sfi_p}\,\rmd x^{\sfi_p}$ reads $\frac{1}{p!}\,A_{\sfi_1\cdots \sfi_p}\,\rmd x^{\sfi_1}\wedge\cdots\wedge \rmd x^{\sfi_p}$ in the standard notation. 
In addition, we define the metric $g^{\sfi_p\sfj_p}$ that corresponds to $g^{\sfi_1\sfk_1}\cdots g^{\sfi_p\sfk_p}\,\delta_{\sfk_1\cdots \sfk_p}^{\sfj_1\cdots \sfj_p}$. 

We do not apply the multiple-index notation to the group index $\cI$\,, but we still use the condensed notation
\begin{align}
 \cA^{\gi\cdots \gj} \equiv \cA^{\gi}\wedge\cdots\wedge \cA^{\gj}\,,\qquad
 \iota_{\gi\cdots \gj} \hat{\kk}^{(p)}\equiv \iota_{\gi}\cdots \iota_{\gj} \hat{\kk}^{(p)}\,.
\end{align}
The square bracket on the group indices, such as $\iota_{[\gi_1\cdots \gi_{q-1}} \hat{\kk}_{\gi_{q}]}^{(p)}$, is used in the usual sense (by including the factorial $\frac{1}{q!}$). 

\section{Generators of the exceptional group and twist matrices}
\label{app:Exceptional}

By following the approach of \cite{hep-th:0307098}, when we consider M-theory, we decompose the generators of $\mathfrak{e}_{\nM(\nM)}$ into the $\mathfrak{gl}(\nM)$ generators $K^i{}_j$, the positive-level generators $\{ R^{i_3},\, R^{i_6},\, \cdots \}$ with upper indices, and the negative-level generators $\{ R_{i_3},\, R_{i_6},\, \cdots \}$ with lower indices. 
The matrix representations $(t_{\bm{\alpha}})^I{}_J$ of these generators in the so-called $R_1$-representation are given by \cite{1111.0459}
\begin{align}
 K^k{}_l &\equiv \tilde{K}^k{}_l - \frac{1}{9-\nM}\,\delta_l^k\,\bm{1} \,, \quad
 \tilde{K}^k{}_l \equiv \begin{pmatrix}
 -\delta_l^i\,\delta_j^k & 0 & 0 & \\
 0 & \delta_{i_2}^{km}\,\delta_{lm}^{j_2} & 0 & \cdots \\
 0 & 0 & \delta_{i_5}^{km_4}\,\delta_{lm_4}^{j_5} & \\
  & \vdots & & \ddots
 \end{pmatrix} ,
\\
 R^{k_3} &\equiv \begin{pmatrix}
 0 & 0 & 0 \\
 \delta_{ji_2}^{k_3} & 0 & 0 & \cdots \\
 0 & -\delta^{j_2 k_3}_{i_5} & 0 \\
 & \vdots & & \ddots \\
 \end{pmatrix} , 
\qquad
 R^{k_6} \equiv \begin{pmatrix}
 0 & 0 & 0 \\
 0 & 0 & 0 & \cdots \\
 \delta_{ji_5}^{k_6} & 0 & 0 \\
 & \vdots & & \ddots \end{pmatrix},\quad \cdots\,.
\end{align}
If we consider the case $\nM>8$, the rank of these matrices is infinite, but we can determine the elements of these matrices successively from the low-level components. 
For our purpose, it is not necessary to know the details of the omitted (high-level) components. 
By using the above generators, we construct the matrix $L=(L^I{}_J)$ as
\begin{align}
 L = \Exp{C_{i_3}\,R^{i_3}}\Exp{C_{i_6}\,R^{i_6}} \cdots\,.
\end{align}
We also construct the matrix $\hat{G}_{IJ}$ by exponentiating the generator $K^k{}_l$ as
\begin{align}
 \hat{G}_{IJ} = \abs{g}^{\frac{1}{9-n}}\begin{pmatrix}
 g_{ij} & 0 & 0 & \\
 0 & g^{i_2j_2} & 0 & \cdots \\
 0 & 0 & g^{i_5j_5} & \\
 & \vdots & & \ddots
 \end{pmatrix}.
\end{align}
Then the generalized metric $\cM_{IJ}$, which has unit determinant, is given by $\cM_{IJ}=(L^{\rmT}\,\hat{G}\,L)_{IJ}$. 

When we study type IIB theory, we decompose the generators of $\mathfrak{e}_{\nM(\nM)}$ into the $\mathfrak{gl}(\nM-1)$ generators $K^p{}_q$, the $\mathfrak{sl}(2)$ generators $R_{\Bc\Bd}$\,, the positive-level generators $\{ R_{\Ba}^{m_2},\, R^{m_4},\, R_{\Ba}^{m_6},\,\cdots \}$ with upper indices, and the negative-level generators $\{ R^{\Ba}_{m_2},\, R_{m_4},\, R^{\Ba}_{m_6},\, \cdots \}$ with lower indices. 
Their matrix representations $(t_{\bm{\alpha}})^I{}_J$ in the $R_1$-representation are given by \cite{1405.7894}
\begin{align}
 K^p{}_q &\equiv \tilde{K}^p{}_q - \frac{1}{9-\nM}\,\delta_q^p\,\bm{1} \,,
\quad
 \tilde{K}^p{}_q \equiv \begin{pmatrix}
 -\delta_q^m\,\delta_n^p & 0 & 0 & 0 & \\
 0 & \delta_{\Ba}^{\Bb}\,\delta_{m}^{p}\,\delta_{q}^{n} & 0 & 0 & \cdots \\
 0 & 0 & \delta_{m_3}^{pr_2}\,\delta_{qr_2}^{n_3} & \\
 0 & 0 & 0 & \delta_{\Ba}^{\Bb}\,\delta_{m_5}^{pr_4}\,\delta_{qr_4}^{n_5} & \\
 & \vdots & & & \ddots
 \end{pmatrix} ,
\\
 R_{\Bc\Bd} &\equiv {\footnotesize {\arraycolsep=0.5mm 
 \begin{pmatrix}
 0 & 0 & 0 & 0 & \\
 0 & \delta^{\Ba}_{(\Bc} \epsilon_{\Bd)\Bb}\delta_m^n & 0 & 0 & \cdots \\
 0 & 0 & 0 & 0 & \\
 0 & 0 & 0 & \delta^{\Ba}_{(\Bc} \epsilon_{\Bd)\Bb}\delta^{n_5}_{m_5} & \\
 & \vdots & & & \ddots 
\end{pmatrix}}},\quad
 R_{\Bc}^{r_2}
 \equiv \begin{pmatrix}
 0 & 0 & 0 & 0 \\
 \delta_{\Bc}^{\Ba}\,\delta_{nm}^{r_2} & 0 & 0 & 0 & \cdots \\
 0 & \epsilon_{\Bc\Bb}\,\delta_{m_3}^{nr_2} & 0 & 0 \\
 0 & 0 & -\delta_{\Bc}^{\Ba}\,\delta_{m_5}^{n_3r_2} & 0 \\
 & \vdots & & & \ddots \\
 \end{pmatrix} ,
\\
 R^{r_4} &\equiv \begin{pmatrix}
 0 & 0 & 0 & 0 \\
 0 & 0 & 0 & 0 & \cdots \\
 \delta^{r_4}_{nm_3} & 0 & 0 & 0 \\
 0 & \delta_{\Bb}^{\Ba}\, \delta^{nr_4}_{m_5} & 0 & 0 \\
 & \vdots & & & \ddots \\
 \end{pmatrix} ,
\qquad
 R_{\Bc}^{r_6} \equiv \begin{pmatrix}
 0 & 0 & 0 & 0 \\
 0 & 0 & 0 & 0 & \cdots \\
 0 & 0 & 0 & 0 \\
 \delta_{\Bc}^{\Ba}\,\delta^{r_6}_{nm_5} & 0 & 0 & 0 \\
 & \vdots & & & \ddots \\
 \end{pmatrix} .
\end{align}
The matrix $L$ in type IIB theory is given by
\begin{align}
 L = \Exp{B^{\Ba}_{m_2}\,R_{\Ba}^{m_2}}\Exp{B_{m_4}\,R^{m_4}}\Exp{B^{\Ba}_{m_6}\,R_{\Ba}^{m_6}}\cdots\,,
\end{align}
and we construct the matrix $\hat{G}_{IJ}$ by using $K^m{}_n$ and $R_{\Ba\Bb}$ as
\begin{align}
 \hat{G}_{IJ} &= \abs{\mathsf{g}}^{\frac{1}{9-n}}{\footnotesize\begin{pmatrix}
 \mathsf{g}_{mn} & 0 & 0 & 0 \\
 0 & m_{\Ba\Bb}\,\mathsf{g}^{mn} & 0 & 0 & \cdots \\
 0 & 0 & \mathsf{g}^{m_3n_3} & 0 \\
 0 & 0 & 0 & m_{\Ba\Bb}\,\mathsf{g}^{m_5n_5} \\
 & \vdots & & & \ddots \\
 \end{pmatrix}},
\quad
 m_{\Ba\Bb} = \Exp{\Phi}\begin{pmatrix} \Exp{-2\Phi} +(C_0)^2 & -C_0 \\ -C_0 & 1 \end{pmatrix}.
\end{align}
Similar to the M-theory case, we then define the generalized metric as $\cM_{IJ}=(L^{\rmT}\,\hat{G}\,L)_{IJ}$. 

The matrix $\cL^I{}_J$ which is introduced in the definition of $\omega^{(F)}_{IJ;\cK}$ in Eq.~\eqref{eq:omega-F} is given by
\begin{alignat}{2}
 \cL &= \Exp{F_{i_3}\,R^{i_3}}\Exp{F_{i_6}\,R^{i_6}} \cdots &&\quad \text{(M-theory)}\,,
\\
 \cL &= \Exp{F^{\Ba}_{m_2}\,R_{\Ba}^{m_2}}\Exp{F_{m_4}\,R^{m_4}}\Exp{F^{\Ba}_{m_6}\,R_{\Ba}^{m_6}}\cdots &&\quad \text{(Type IIB theory)}\,.
\end{alignat}
This shows that in our approach, the field strengths $F_{p+1}\equiv \rmd A_{p}$ of the worldvolume gauge fields are in one-to-one correspondence with the positive-level generators $R^{i_{p+1}}$. 

By using an invariant tensor $\eta_{IJ;\cK}$, known as the $\eta$-symbol, that connects a symmetric product $R_1\otimes_S R_1$ and the $R_2$-representation, we can define the $\mathfrak{e}_{\nM(\nM)}$ generators in the $R_2$-representation, denoted by $(t_{\bm{\alpha}})^\cI{}_\cJ$, through
\begin{align}
 (t_{\bm{\alpha}})^L{}_I\, \eta_{LJ;\cK} + (t_{\bm{\alpha}})^L{}_J\, \eta_{IL;\cK} = \eta_{IJ;\cL}\,(t_{\bm{\alpha}})^\cL{}_\cK \,.
\end{align}
In the M-theory parameterization, we find
\begin{align}
 (R^{k_3})^{\cI}{}_{\cJ} \equiv 
 \begin{pmatrix}
 0 & -\delta^{ik_3}_{j_4} & \cdots \\
 0 & 0 \\
 \vdots & & \ddots 
\end{pmatrix} , \qquad
 (R^{k_6})^{\cI}{}_{\cJ} \equiv 
 \begin{pmatrix}
 0 & 0 & \cdots \\
 0 & 0 \\
 \vdots & & \ddots 
\end{pmatrix},\quad \cdots \,,
\end{align}
and in the type IIB parameterization, we have
\begin{align}
 &(R^{p_2}_{\Bc})^{\cI}{}_{\cJ} \equiv 
 \begin{pmatrix}
 0 & -\epsilon_{\Bb\Bc}\,\delta^{p_2}_{n_2} & 0 \\
 0 & 0 & -\delta^{p_2m_2}_{n_4} & \cdots \\
 0 & 0 & 0 \\
 & \vdots & & \ddots 
\end{pmatrix} , \qquad
 (R^{p_4})^{\cI}{}_{\cJ} \equiv 
 \begin{pmatrix}
 0 & 0 & \delta^{\Bb}_{\Ba}\,\delta^{p_4}_{n_4} \\
 0 & 0 & 0 & \cdots \\
 0 & 0 & 0 
\\
 & \vdots & & \ddots 
\end{pmatrix},
\\
 &(R^{p_6}_\gamma)^{\cI}{}_{\cJ} \equiv 
 \begin{pmatrix}
 0 & 0 & 0 \\
 0 & 0 & 0 & \cdots \\
 0 & 0 & 0 \\
 & \vdots & & \ddots 
\end{pmatrix} , \quad \cdots \,.
\end{align}
By using these, the matrix $\cL^{\cI}{}_{\cJ}$ appearing in Eq.~\eqref{eq:omega-F} can be defined as
\begin{alignat}{2}
 \bigl(\cL^{\cI}{}_{\cJ}\bigr)&= \Exp{F_{i_3}\,R^{i_3}}\Exp{F_{i_6}\,R^{i_6}} \cdots&&\quad \text{(M-theory)}\,,
\\
 \bigl(\cL^{\cI}{}_{\cJ}\bigr)&= \Exp{F^{\Ba}_{m_2}\,R_{\Ba}^{m_2}}\Exp{F_{m_4}\,R^{m_4}}\Exp{F^{\Ba}_{m_6}\,R_{\Ba}^{m_6}} \cdots&&\quad \text{(Type IIB theory)}\,,
\end{alignat}

\end{document}